\documentclass[12pt,leqno]{amsbook}
 
\usepackage{amsmath}
\usepackage{graphicx}
\usepackage{color}
\input amssym.def
\input amssym

\long\def\symbolfootnote[#1]#2{\begingroup%
\def\thefootnote{\fnsymbol{footnote}}\footnote[#1]{#2}\endgroup}

\newtheorem{theorem}{\sc Theorem}[section] 
\newtheorem{lemma}[theorem]{\noindent {\sc Lemma}} 
\newtheorem{corollary}[theorem]{\sc Corollary}
\newtheorem{proposition}[theorem]{\sc Proposition}

\newtheorem{definition}{Definition}[theorem]
\newtheorem{conjecture}{Conjecture}[theorem]

\theoremstyle{plain}

\renewcommand{\a}{\alpha}
\renewcommand{\b}{\beta}
\renewcommand{\d}{\delta}

\newcommand{\e}{\varepsilon}
\renewcommand{\th}{\theta}
\newcommand{\g}{\gamma}
\newcommand{\G}{\Gamma}
\renewcommand{\l}{\lambda}
\renewcommand{\k}{\kappa}
\renewcommand{\L}{\Lambda}

\newcommand{\s}{\sigma}

\renewcommand{\t}{\tau}
\newcommand{\cal}{\mathcal}
\newcommand{\Z}{{\Bbb Z}}
\newcommand{\R}{{\Bbb R}}
\newcommand{\C}{{\Bbb C}}

\renewcommand{\o}{\omega}

\renewcommand{\i}{\infty}
\newcommand{\p}{\partial}

\renewcommand{\thefootnote}{\fnsymbol{footnote}}
\newcommand{\nat}{\natural}
\renewcommand{\thefootnote}{\fnsymbol{footnote}}
\addtocontents{toc}{\hfill Page\endgraf}
\addtocontents{lof}{\ Figure\hfill Page}

\makeatletter
\renewcommand\tableofcontents{%
  \newfont{\scaledfont}{cmr12 scaled 7000}
  \parindent\z@\raggedright
  \scaledfont\contentsname\par\normalsize%
  \rule{\textwidth}{1pt}
  \nobreak
  \vskip 40\p@
  \@starttoc{toc}%
}
\makeatother
 \author[J. Harrison  Department of Mathematics  U.C. Berkeley]{Jenny Harrison
\\Department of Mathematics
\\University of California, Berkeley\\December 31, 2004}
\title[Ravello Lecture Notes]{Ravello Lectures on Geometric Calculus -- Part I}
\email{harrison\@@math.berkeley.edu}
\date{December 31, 2004}
\begin{document}

\maketitle

 \chapter*{Preface}   
 
In these notes we present a new approach to calculus in which more efficient choices of limits are taken at key points of the development. For example, $k$-dimensional tangent spaces are replaced by representations of simple $k$-vectors supported in a point as limits of simplicial $k$-chains in a Banach space (much like Dirac monopoles).  This subtle difference has powerful advantages that will be explored.  Through these ``infinitesimals'', we obtain a coordinate free theory on manifolds that builds upon the Cartan exterior calculus. 
An infinite array of approximating theories to the calculus of Newton and Lebiniz becomes available and we can now revisit old philosophical questions such as  which models are most natural for the continuum or for physics. 

   Within this new theory of Geometric Calculus are found the classical theory on smooth manifolds, as well as three distinct, new extensions.  All three can be seen as part of the space of polyhedral chains completed with respect to what the author calls the ``natural norm''.   The author calls elements of the Banach space obtained upon completion  ``chainlets''.     
    
 We take many viewpoints in  mathematics and its applications, be it smooth manifolds, Lipschitz structures, polyhedra, fractals, finite elements, soap films, measures, numerical  methods, etc.  The choice sets the stage and determines our audience and our methods.      No particular viewpoint is ``right'' for all applications. The Banach space of chainlets unifies these viewpoints.

 \begin{enumerate}
 \item {\bf Discrete Calculus}\symbolfootnote[1]{Some aspects of this are contained in the paper \cite{hodge} and slides prepared for the Caltech meeting in October, 2003 on Discrete Geometry for Mechanics are available on line at http://math.berkeley.edu/$\sim$harrison.} with convergence to the smooth continnuum, including more general theorems of Stokes, Green and Gauss. 
 \item {\bf   Bilayer Calculus} with applications to the calculus of variations including Plateau's problem (soap bubbles).
  \item { \bf Calculus on Fractals}\end{enumerate}

The discrete theory is perhaps the most far reaching.  Poincar\'e taught us to use a simplicial complex as the basic discrete model.  It has become fashionable to use Whitney forms \cite{whitney} as cochains and they are based on the simplicial complex.   However, there is a great deal information within a simplicial complex that is not needed for calculus.  There are corners, matching boundaries of simplices, ratios of length to area, and so forth.   The standard proof to Stokes' theorem relies on boundaries matching and cancelling with opposite orientation where they meet.  However,  problems of approximation sometimes arise in the continuum limit, as a finer and finer mesh size is used.   For example, 
\begin{itemize}
\item Supposed approximations may not converge
\item Vectors, especially normal vectors, may be hard to define
\item 	Cochains fail to satisfy a basic property such as  commutativity or associativity of wedge product, or existence of Hodge star.
\end{itemize}

   One may discard much of the information in a simplicial complex and instead use sums of what are essentially ``weighted oriented monopoles and higher order dipoles'' as approximators for both domains and integrands.   (These will be fully described in the lectures.)  Standard operators such as boundary, coboundary, exterior derivative and Laplace are naturally defined on these ``element chains''. Integral relations  of calculus are valid and essentially trivial at this discrete level and converge to the continuum limit.  Vectors have discrete counterparts, including normal vectors, and discrete cochains are well behaved.  
   
   One of our goals in these notes is to write the theorems of Stokes, Gauss and Green in such a concise and clear manner that all manner of domains are permitted without further effort.    We obtain this  by first treating the integral as a bilinear functional on pairs of integrands and domains, i.e., forms and chainlets.  Secondly, we define a geometric Hodge star operator $\star$ on domains $A$ and proving for $k$-forms $\o$ in $n$-space
   $$\int_{\p A} \o = \int_A d\o$$ and
   $$\int_{\star A} \star \o =\int_A \o$$ with appropriate assumptions on the domain and integrand (see Chapter \ref{fractals}).  
   One may immediately deduce
   $$\int_{\star \p A} \o =  (-1)^{k(n-k)}\int_A d \star \o$$
   and 
   $$\int_{\p \star A} \o = (-1)^{(k+1)(n-k-1)} \int_A \star d\o$$ which are generalized and optimal versions of the divergence and curl theorems, respectively.

   Arising from this theory are new, discrete approximations for the real continuum. Morris Hirsch wrote,\symbolfootnote[2]{e-mail message, quoted with permission}     
 on December 13, 2003.  \begin{quotation}  A
basic philosophical problem has been to make sense of ``continuum'',
as
in the space
of real numbers, without introducing numbers.  Weyl \cite{weyl} wrote,   ``The
introduction of coordinates as numbers... is an act of violence''.
Poincar\'e wrote about the ``physical continuum'' of our intuition, as
opposed
to the mathematical continuum.  Whitehead (the philosopher) based our
use of
real numbers on our intuition of time intervals and spatial regions.
The
Greeks tried, but didn't get very far in doing geometry without real
numbers.
But no one, least of all the Intuitionists, has come up with even a
slightly
satisfactory replacement for basing the continuum on the real number
system,
or basing the real numbers on Dedekind cuts, completion of the
rationals, or
some equivalent construction.

Harrison's theory of chainlets can be viewed as a different way to build topology out
of
numbers.  It is a much more sophisticated way, in that it is (being)
designed
with the knowledge of  what we have found to be geometrically useful
(Hodge
star, Stokes' theorem, all of algebraic topology,. . .), whereas the
standard
development is just ad hoc-- starting from Greek geometry, through
Newton's
philosophically incoherent calculus, Descarte's identification of
algebra with
geometry, with additions of abstract set theory, Cauchy sequences,
mathematical logic, categories, topoi, probability theory,  and
so
forth, as needed.  We could add quantum mechanics, Feynman diagrams
and
string theory! The point is this is a very roundabout way of starting
from
geometry, building all that algebraic machinery, and using it for
geometry
and physics.  I don't think chainlets, or any other purely mathematical
theory,
will resolve this mess, but it might lead to a huge simplification of
important parts of it.  
 
\end{quotation}

Readers may choose any number of points of view.  Smooth manifolds, discrete element chains, polyedra, fractals, any dense set of chainlets will produce the same results in the limit, which links together numerous approaches some of which seemed distantly related, at best.  Philosophically, we may not know the most natural  models for physics, but we provide here various options which all converge to the continuum limit and are consistent with each  other and standard operators of mathematics and physics.   We anticipate applications to physics, continuum mechanics, biology, electromagnetism, finite element method, PDE's, dynamics, computation, wavelets, vision modeling as well as to pure mathematics (topology, foundations, geometry, dynamical systems)   and give examples in the notes that follow.  Indeed, the author proposes that the discrete theory provides a foundation for new models for quantum field theory, a topic under development.  Algebraic/geometric features of the models make them especially enticing.  It is worse to be   oblivious to the importance of a new theory than to be overly excited, and the author chooses to err on the side of the latter.  
 
I am grateful to the Scientific
Council of GNFM for inviting me to give this course, and to the Director of
the Ravello Summer School, Professor Salvatore Rionero, for his elegant
hospitality.  I also wish to thank Antonio Di Carlo,  
Paolo Podio-Guidugli, Gianfranco Capriz and the other participants of the Ravello Summer School for their interest in my work and for
encouraging me to put my scribbled lecture notes into their present form.  Above all, I thank Morris Hirsch and James Yorke for listening over these  years.  Their support and encouragement have been critical to the success of this research program.   
\begin{itemize}
\item  Harrison, Jenny, {\em  Stokes'
 theorem on nonsmooth chains}, Bulletin AMS, October
 1993.
\item  --, {\em Continuity of the Integral as a Function of the Domain}, Journal of Geometric Analysis, 8 (1998), no. 5, 769--795.
\item  --.Isomorphisms differential forms and cochains, Journal of Geometric Analysis, 8
(1998), no. 5, 797--807.
 \item  --, {\em  Geometric
 realizations of currents and distributions},  Proceedings of Fractals and Stochastics III, Friedrichsroda, German, 2004.
\item  --,{\em Geometric Hodge star operator with applications to the theorems of Gauss and Green}, to appear, Proc Cam Phil Soc.
\item --, {\em Cartan's Magic Formula and Soap Film structures}, Journal of Geometric Analysis
\item --, {\em On Plateau's Problem with a Bound on Energy}, Journal of Geometric Analysis
\item --, {\em Discrete Exterior Calculus with Convergence to the Smooth Continuum}, preprint in preparation 
\item--, {\em Measure and dimension of chainlet domains}, preprint in preparation

\end{itemize}
\newpage
\section*{Geometric Integration Theory}

 {\em  Geometric Integration Theory} (GIT), the great classic of Hassler Whitney \cite{whitney}, was an attempt to articulate the approach to  calculus favored by Leibnitz of approximation of domains by polyhedral chains rather than by parametrization by locally smooth coordinate charts.   
 
 Approximation by polyhedral chains defined using finite sums of cells, simplexes, cubes, etc., has been a technique commonly used by applied mathematicians, engineers,  physicists and computer scientists but has not had the benefit of rigorous mathematical support to guarantee convergence.    We often create grids around a domain depending on a small scale parameter, perform calculations on each grid, let the parameter tend to zero, and hope the answers converge to something meaningful.    Sometimes the limit appears to exist, but actually does not.   An example, given by Schwarz,   is that of a cylinder with unit area, but approximated by simplicial chains with vertices on the cylinder whose areas limit to any positive number in the extended reals.

 Whitney's theory was based on the idea of completing the space of polyhedral chains with a norm and proving continuity of basic operators of calculus with respect to that norm, thereby obtaining a theory of calculus on all points in the Banach space.   This approach had worked well for spaces of functions and Whitney was the first to try it out for domains of integration.  
 
 Unfortunately, his definitions from the late 1940's brought with them serious technical obstructions that precluded extension of most theorems of calculus.   His first norm, the {\itshape \bfseries  sharp norm}, does not have a continuous boundary operator so one cannot state the generalized Stokes' theorem  $$\int_{\p A} \o = \int_A d \o$$ for elements $A$ of the Banach space obtained upon completion.
(It is not necessary for the reader to know the definition of the sharp norm as given by Whitney to understand these notes.)
 
 His second norm, the {\itshape \bfseries flat} norm, has a continuous boundary operator built into it, but does not have a continuous Hodge star operator, and thus one may not state either a generalized divergence or curl theorem in the flat normed space.  (See Table 1.) In this introduction, we use terminology such as ``flat norm'' and ``Hodge star operator'' that will be defined in later sections.  
 \newpage \section*{Geometric Measure Theory}
  Geometric measure theory \cite{federer}, the study of domains through weak convergence and measures,  took the approach of using dual spaces of differential forms and had greater success in extending calculus.   The extension of the Gauss-Green theorem, credited to de Giorgi and Federer,  was a striking application of GMT. Their divergence theorem holds for an $n$-dimensional  current $C$
with $\cal{L}^n$ Lebesgue measurable support and $\cal{H}^{n-1}$  Hausdorff measurable current boundary.  The
vector field $F$ is assumed to be Lipschitz.  Their conclusion takes the form $$\int_{\p C} F(x) \cdot n(C,x)d\cal{H}^{n-1} x = \int_C div F(x) d\cal{L}^n x.$$
The hypotheses imply the existence a.e. of measure theoretic normals
n(C, x) to the (``current'') boundary.  See \cite{federer}

A second crowning achievement of GMT was a solution to the problem of Plateau.   Plateau  \cite{plateau} studied soap films experimentally and noticed that soap film surfaces could meet in curves in only two ways:    Three films can come together along a
 curve at equal 120 degree angles, and four such curves can meet at a point at equal angles of about 109 degrees.  

The question arose:  Given a closed loop of wire, is there a surface with minimal area spanning the curve?  Douglas won the first Field's medal for his solution in the class of smooth images of disks.   Federer and Fleming were able to provide a solution within the class of embedded, orientable surfaces.     They do not model branched surfaces and do not permit the Moebius strip solution as a possible answer, though such solutions do occur in nature.   

The most recent area of application of GMT has been with fractals and that has reinervated interest in the theory   as it provides rigorous methods for studying integral and measure of  nonsmooth domains.   It is generally accepted that nonsmooth sets appear commonly in mathematics as well as in the real world.  Paths of fractional Brownian motion  have become fashionable as models of everyday phenomenon such as stock market models, resulting in a growing demand for results  of geometric measure theory.

  \section*{Geometric Calculus}
  In these lectures will be developed a third approach that the author calls ``geometric calculus'' (GC).      The theory of GC begins  in Lecture 1 begins with a norm   the author calls the ``natural norm''\symbolfootnote[3]{This is a pun on Whitney's musically inspired terms, sharp and flat.}   on the linear space of polyhedral $k$-chains.  In contrast to Whitney's norms the boundary and star operators are  bounded.  (See Table 1.)  It follows that they are bounded on the completion $\cal{N}_k^{\i}$ of this normed space which the author calls the space of ``$k$-chainlets''.   Similarly, the integral is bounded on polyhedral $k$-chains in the natural norm.  Lecture 2 presents necessary background on exterior calculus.  In Lecture 3 we prove a de Rham theorem for  dual spaces of chainlets and differential forms.  In Lecture 4 we present calculus on chainlets.  For example, proving Stokes' theorem on polyhedral chains or on any other dense set in $\cal{N}^{\i}$ implies it holds on the entire space.      We  prove a general divergence theorem for a large class of domains.     Thus we will able to measure flux across surfaces such as the lungs.  (There should be broad application in applied mathematics for this result.)   The generalized divergence theorem is valid for chainlets of any codimension with boundaries that are not even locally finite or locally Euclidean.  There is no assumption on the existence of normal vectors anywhere, or even the existence of measure theoretic normals.  Our result applies to all currents satisfying the hypotheses of the result of Federer and de Giorgi. 

  \begin{table}\caption{Bounded operators on norms}\begin{tabular}{|c|c|c|c|c|}\hline norm & d & $\star$ & Stokes' theorem & Divergence theorem \\\hline sharp & no & yes  & no & no\\\hline flat & yes & no  & yes & no\\\hline natural & yes & yes & yes & yes\\\hline \end{tabular}
\end{table}
  
 The theory of GC  has been equally motivated by both GIT and GMT.    
   
 A legitimate question is whether this approach is at least as powerful and far reaching.  These lectures will examine the three major applications of GMT mentioned above, and  will show in what manner each can be improved and simplified. 
  We give additional applications of discrete, fractal and bilayer calculus that have not been significantly developed in GMT or GIT.    That GC is simpler than GMT in both concept and execution is a side benefit.     We cannot overemphasize the importance of bringing algebra to the domain of integration, as well as the integrand, making the integral a bilinear pairing on forms and chainlets.

As noted above, any dense subset of chainlets leads to a full theory of calculus in the limit.  However, different choices of subsets may present quite different initial results before taking limits in the chainlet space.  For example, it is shown in Lecture 5 that smooth differential forms are dense in the space of chainlets, as are smooth submanifolds.   Lecture 6 presents a theory of locally compact abelian groups leading to a theory of distributions with the notion of differentiaion.    Lecture 7  introduces a dense subset of newly discovered chainlets  consisting of finite sums of what the author calls ``$k$-elements''.  Each   $k$-element is supported in a single point and makes rigorous   the notion of an ``infinitesimal'' of Newton and Leibnitz.     We distinguish two meanings of an ``infinitesimal'', although we do not use this term beyond the introduction.  One is an ``infinitesimal domain'', the other is its dual, an ``infinitesimal form''.  A  $k$-element is a kind of infinitesimal domain.  

Higher order   $k$-elements of order $s$  are also introduced in Lecture 7.     A $k$-element of order zero   is the same as a   $k$-element which can be thought of as a geometric analogue of the Dirac delta point or a monopole. In the natural norm Banach space the distance between two monopoles at points $p$ and $q$ is proportional to the distance between $p$ and $q$.  Contrast this with the theory of distributions where metrics are defined weakly.   Each  $k$-element of order $s$  is again supported in a point.     For $s = 1$ we have a dipole, $s = 2$, a quadrupole, and so forth.  The number $s$ corresponds loosely to an s order derivative of a delta function, though of course our elements are delta points, the geometric analogs of delta functions and are defined in arbitrary dimension and codimension.  Ê There is no upper bound on $s$, even in dimension one, although $k$ is bounded by the ambient dimension.  Each  $k$-element of order $s$ has a well-defined boundary as a sum of  $(k-1)$-element or order $(s+1)$ and  it has a discrete analog of a normal bundle defined via the geometric star operator.   It is striking that the divergence theorem can be proved at a single point.  The space $\cal{ N}^{\i}$  extends standard geometric structures by adding geometric and algebraic attributes to points.         We form our basic ``multiorder $k$-element chains'' from the direct sum of spaces of $k$-element chains of order $s$.   The direct sum is required for the boundary and star operators to be closed.    Since such ``discrete'' chains are dense in the space of chainlets, we may develop a discrete approach to the full calculus. Elements of the dual space to $k$-element chains    converge to smooth  differential forms in the limit.    With this we are able to define operators on forms as dual to geometrically defined operators on $k$-elements.  The forms, which ``measure'' cells, become perfectly matched for their job, in a coordinate free fashion.     In the full discrete theory we quantize matter and energy with models that make their identification transparent, up to a constant.\footnote{$e = m c^2$}  
 
It is a subtle, but significant difference to begin with $k$-element chains (GC) rather than polyhedral chains (GIT) or $k$-vectors (GMT).  With our approach, at the starting point of GC, we reveal a common denominator for forms and chains that is the essence of Poincar\'e duality.   We make this precise via an extension of tensor analysis, using  what the author calls ``multitensors'' that are monopoles, dipoles, quadrupoles, etc.  It is striking that the subspace of $k$-element chains is reflexive, although the space of chainlets is not, neither are other subspaces such as differential forms or polyhedral chains.     In the application to smooth manifolds, there is no longer reliance on tangent spaces, the hardy tools of mathematics.  Instead we use the versatile and rich $k$-elements which have boundaries defined, where other operators apply and the essence of calculus is found.  Linear algebra is valid on $k$-elements and carries over locally to chainlets, much as it has done for manifolds via tangent spaces.  The use of $k$-elements in place of $k$-dimensional tangent spaces opens up new vistas well beyond the scope of the integral methods of GIT and that is why we prefer to call it GC.  New fields of enquiry emerge which will be highlighted as the lectures develop.

 {\bf Outline}
\begin{enumerate}
\item[1.] Chainlets 
\item[2.] Exterior algebra
\item[3.] Differential forms and cochains (de Rham theorem)
\item[4.]  Calculus on fractals -- star operator (no assumptions of self similarity, connectivity, local Euclidean, or locally finite mass.  Chainlets are useful for modeling structures in the physical sciences with nontrivial local structures.)
\item[5.] Poincar\'e duality for chains and cochainlets (chainlet representations of differential forms, cap product, wedge product, convolution product)
\item[6.] Locally compact abelian groups
\item[7.] Discrete calculus (part I) (a coordinate free approach that contains calculus on smooth manifolds, numerical methods, and more)  The first half of this chapter appears in the first part of these lecture notes presented below.  The second half, as well as the following chapters, will appear in the second part to follow.  
\item[8.]  Bilayer calculus (soap film structures, calculus of variations on soap films, Lipid bilayers, immiscible fluids, fractures, multilayer calculus, bilayer foliations, Frobenius theorem, differential equations)
\item[9.]  Measure theory (lower semi continuity of $s$-mass, $k$-vector valued additive set functions, $s$-normable chainlets)
\item[10.] Applications
\end{enumerate}

Because these notes are  only meant as an introduction to the theory, a number of important results, of necessity, have been left out.    

\chapter{Chainlets}\label{chainlets}
\renewcommand{\thetheorem}{1.\arabic{theorem}}
\renewcommand{\thesection}{1.\arabic{section}}
\setcounter{theorem}{0}
\setcounter{section}{0}

\section*{Polyhedral chains}
 A {\itshape \bfseries cell} $\s$  in $R^n$ is the nonempty intersection of finitely many closed 
affine half spaces.   The {\itshape \bfseries dimension} of $\s$ is $k$  if $k$ is the dimension of the smallest  affine  
subspace $E$ containing
$\s$.  The {\itshape \bfseries support} $spt \sigma$ of  $\sigma$ is the set of all points  in the intersection of half spaces that determine $\sigma$.   

Assume $k >0$.  An {\itshape \bfseries orientation} of $E$ is an equivalence class of ordered bases for the linear subspace 
parallel to $E$ where two bases are equivalent if and only if their transformation matrix has positive determinant.  An {\itshape \bfseries orientation} of $\s$ is defined to be an orientation of its subspace $E$.     Henceforth, all $k$-cells are assumed to be oriented.    (No orientation need be assigned to 
$0$-cells which turn out to be single points $\{x\}$ in $\R^n$. )   When $\s$ is a simplex,  each orientation determines an equivalence class of orderings of the set of vertices of $\s$.  

An {\itshape \bfseries algebraic $k$-chain} is a (formal) linear combination of oriented $k$-cells with coefficients in $G =
\Z$ or
$\R$. The vector space 
of algebraic $k$-chains is the quotient of the vector space generated by 
oriented $k$-cells by the subspace generated by chains of the form $\s + \s'$  where 
$\s'$ is obtained from $\s$ by reversing its orientation.  

    Let $P= \sum a_i \sigma_i$ be an algebraic $k$-chain.   Following Whitney \cite{whitney} define the function $P(x) := \sum a_i$ where the sum is taken over all $i$ such that $x \in spt \sigma_i.$   Set  $P(x) := 0$ if $x$ is not in the support of any $\sigma_i$.
We say that algebraic $k$-chains $P$ and $Q$ are {\em equivalent} and write $P \sim Q$ iff the functions $P(x)$ and $Q(x)$ are equal except in a finite set of cells of dimension $< k.$   For example, $(-1,1) \sim (-1,0) + (0,1).$  A  {\em polyhedral $k$-chain} is defined as an equivalence class of 
algebraic $k$-chains.
 This clever definition implies  that if $P'$ is a subdivision of the algebraic chain $P$, then $P$ and 
$P'$ determine the same polyhedral chain which behaves nicely for integrating 
forms. In particular, algebraic $k$-chains $P$ and $Q$ are equivalent iff integrals of smooth differential $k$-forms agree over them.  This property is sometimes taken as the definition of a polyhedral chain, but we wish to define it without reference to differential forms. 
  If $P$ is an algebraic chain $[P]$ denotes the polyhedral chain of $P$.   As an abuse of notation we usually omit the square brackets and write $P$ instead of $[P]$.
Denote the linear space of polyhedral chains by $\cal{P}_k.$

 \noindent{\bf Remarks}.  Every polyhedral chain $P$ has a nonoverlapping representative.  Two cells that have the same coefficient, but opposite orientation will cancel each other where they overlap.   If they have the same orientation, their coefficients are added.    

The standard boundary operator $\p$ on $k$-cells $\sigma$ produces an algebraic $(k-1)$-chain.  This extends linearly to a boundary operator on algebraic $k$-chains.  This, in turn, leads naturally to a well defined boundary operator $\p$ on polyhedral  $k$-chains  for
$k \ge 1$. For $k = 0 $ we set $\p P := 0$.
\[P_n \buildrel \p \over \to P_{n-1}
 \buildrel \p \over \to   \cdots   \buildrel \p \over \to P_1 \buildrel \p \over \to P_0\]
is a chain complex since $\p\circ \p = 0.$  (We omit the proof which is standard.)

\subsection*{Mass of polyhedral chains} \quad \\ Let $M(\sigma)$ denote k-dimensional Lebesgue measure, or k-volume of a $k$-cell
$\sigma$.   Every 0-cell $\s^0$ takes the form $\s^0 = \{x\}$ and we set $M(\s^0) = 1$.
  The {\itshape \bfseries mass} of $P$ is defined by
\[M(P) := \sum_{i=1}^m
|a_i|M(\sigma_i)\]
where  $P = \sum_{i=1}^m a_i\sigma_i$ and the cells $\sigma_i$ are non-overlapping.  We think of mass as weighted $k$-volume.  For example, the mass of a piecewise linear curve with multiplicity two is twice its arc length. Mass is a norm on the vector
space
$\cal{P}_k$.  Suppose $ \sum_{i=1}^m a_i \sigma_i$ is a non-overlapping representative of $P$.    The {\itshape \bfseries support} of $P$
is defined as 
$spt(P) := \cup  
spt(\sigma_i).$   

It is worth noting to those well versed in analysis based on  unions and intersections of sets that these definitions are substantially different and bring algebra of multiplicity and orientation into the mathematics at an early stage.

 The norms are initially defined for polyhedral $k$-chains in Euclidean space $\R^n$ and it is shown later how to extend the results to singular $k$-chains in  Riemannian manifolds $M^n$.

 \subsection*{Difference cells}

For $v \in \R^n$ let $|v|$ denote its norm and $T_v$ translation through $v$.  Let   $\s^0$ be a $k$-cell in $\R^n$.  For consistency of terminology we also call $\s^0$ a {\itshape \bfseries $0$-difference $k$-cell.} Let $v_1 \in \R^n$.  Define the {\itshape \bfseries $1$-difference $k$-cell} $$\s^1 := \s^0 - T_{v_1}\s^0.$$   
This is a chain consisting of two cells, oppositely oriented.  A simple example consists of the sum of the opposite faces of a cube, oppositely oriented.  The chain is supported in these two faces. Since it is not always true that $T_v\s + T_w\s = T_{v+w} \s$ one has to take care with adding two multicells:  $\s - T_v \s + \s - T_w \s \ne 2\s - T_{v+w}\s.$    Given $\s^0$ and $v_1, \cdots, v_r \in \R^n$, define the {\itshape \bfseries $j$-difference $k$-cell} inductively \[\s^{j+1} := \s^j  - T_{v_{j+1}} \s^j.\]  The integer $k$ is the {\em dimension} of the multicell, $j$ is its {\itshape \bfseries order}.  \footnote{This terminology is distinguished from the $j$-differential $k$-cells and $j$-differential $k$-chainlets that we will define later which are geometric directional derivatives, and thus come from infinitesimal translations.}

 A {\itshape \bfseries $j$-difference $k$-cellular chain} $D^j$  in $\R^n$ is a (formal) sum of $j$-difference $k$-cells,
$$D^j =
\sum_{i=1}^m a_i \s_i^j$$
with coefficients $a_i \in G.$    The vector space   $\cal{D}_k^j$
of $j$-difference $k$-cellular chains is the quotient of the vector space generated by $j$-difference $k$-cells 
  by the subspace generated by multicellular chains of the form $(\s - T_v \s) + (\s' -T_v \s')$  where 
$\s'$ is obtained from $\s$ by reversing its orientation.

\subsection*{Difference norms} Given a $j$-difference $k$-cell $\s^j$ in $\R^n$ generated by a $k$-cell $\s^0$ and vectors $v_1,  \cdots , v_j$, define $\|\s^0\|_0 := M(\s^0)$ and for $j \ge 1$, 
$$\|\s^j\|_j := M(\s^0)|v_1| |v_2| \cdots |v_j |.$$
 For $D^j = \sum_{i=1}^m a_i \s_i^j$, possibly overlapping, define its {\itshape \bfseries difference norm} as $$\|D^j\|_j := \sum_{i=1}^m |a_i|\| \s_i^j\|_j.$$


\subsection*{$r$-natural norms}
  Let $P \in \cal{P}_k$ be a polyhedral $k$-chain. For $r = 0$ define
$$|P|^{\natural_0} := M(P).$$
For $r \ge 1$ define the {\itshape \bfseries r-natural} norm
$$|P|^{\natural_r} : = \inf\left\{\sum_{j=0}^r\|D^j\|_j + |C|^{\natural_{r-1}} \right\}$$  
where the infimum is taken over all decompositions
$$P = \sum_{j=0}^r D^j + \p C$$
where $D^j \in \cal{D}_k^j$ and $C  \in \cal{P}_{k+1}.$  There is at least one decomposition, a trivial one, where $D^0 = P$ and where $C$ and the other $D^j$ are zero chains.
   It is clear $|\quad|^{\natural_r}$ is a semi-norm. 
 We prove it is a norm in Chapter \ref{forms}.  This norm is often quite difficult to compute. We are more interested in which sequences are Cauchy sequences in this norm.

It follows immediately from the definitions that the boundary operator
on chains is bounded w.r.t. the r-natural norms.
\begin{proposition}\label{lem.boundary} If $P \in  \cal{P}_k$ then
$$|\p P|^{\natural_{r}} \le  |P|^{\natural_{r-1}}.$$
\end{proposition}

\medskip \noindent {\bf Exercise 1}  Find a Cauchy sequence of polyhedral chains  in the $1$-natural norm that converges pointwise to the graph of the Weierstrass nowhere differentiable function $y = \sum 2^{-k} sin(2^{3k}x).$ 
 
 
\medskip \noindent {\bf Exercise 2} In the plane, define a sequence of polyhedral chains $P_k$ as follows:   Let $\s_k$ denote the positively oriented square centered at the origin with edge $2^{-k}.$  Let $P_k = 2^{2k} \s_k.$   Prove that the $P_k$ form a Cauchy sequence in the $1$-natural norm.  The boundaries $\p P_k$ form a Cauchy sequence in the $2$-natural norm.

  \setcounter{chapter}{1} 
\chapter{ Exterior algebra}\label{algebra}
\renewcommand{\thetheorem}{2.\arabic{theorem}}
\renewcommand{\thesection}{2.\arabic{section}}
\setcounter{theorem}{0}
\setcounter{section}{0}

\section*{$k$-vectors}   The idea of a vector in $\R^n$ as an arrow with length and direction extends to higher dimensions.   For $0 \le k \le n$,  we may think of a ``simple $k$-vector''  in $\R^n$ much like a small parallelopiped, with well defined $k$-volume, contained in a $k$-dimensional subspace of $\R^n$, called a ``$k$-direction''.    We present two approaches for making this precise.  The first is well established and comes from the exterior calculus of Cartan.  See \cite{flanders}, for example.  The second approach will be presented in Chapter \ref{discrete}.  

Let $\R$ denote the field of reals with elements denoted $a,b,c, \dots$ and 
$V$ an $n$-dimensional vector space over  $\R$ with elements denoted $\a,\b,\g, \dots$.  For each $k = 0,1,2, \dots$, we construct a new vector space $$\Lambda^kV$$ over $\R$ called the {\itshape \bfseries space of $k$-vectors in $V$}.  By definition
$$\Lambda^0V:= \R$$
$$\Lambda^1V:= V.$$
 
The space $\Lambda^2 V$ of $2$-vectors of $V$ consists of all formal sums $$\sum a_i (\a_i \wedge \b_i)$$ modulo the subspace generated by relations
$$(a\a) \wedge \b = a(\a \wedge \b) = \a \wedge(a \b),$$
$$(\a_1 + \a_2) \wedge \b = (\a_1 \wedge \b) + (\a_2 \wedge \b),$$ and $$ \a \wedge \b = -\b \wedge \a.$$ In particular, $\a \wedge \a = 0.$ We call $\a \wedge \b$  the {\itshape \bfseries exterior product} of $\a$ and $\b$.   A {\itshape \bfseries  simple $2$-vector} is a $2$-vector that can be written as  $\a \wedge \b.$

\medskip \noindent{\bf Exercise}:  $\a$ and $\b$ are dependent iff $\a \wedge \b = 0.$

\medskip \noindent {\bf Exercise}:   \medskip Suppose $\s^1 , \cdots , \s^n$ is a basis of $V$.   Prove that $\s^i \wedge \s^j, i < j$, forms a basis of $\Lambda^2V.$  Hence $dim \Lambda^2V  =
\left(\begin{array}{c}n \\2 \end{array}\right)$

(See \cite{flanders}, for example, for details)
The space $\Lambda^k(V)$ of $k$-vectors of $V$, $2  < k \le n$, is defined recursively.  A {\itshape \bfseries $k$-vector} is  a formal sum $$\sum a_i \a_i \wedge \b_i$$ where $\a_i \in \Lambda^{k-1}(V)$ is a simple $(k-1)$-vector and $\b_i \in V$, subject to the constraints $$(a\a) \wedge \b = a(\a \wedge \b) = \a \wedge(a \b),$$
$$(\a_1 + \a_2) \wedge \b = (\a_1 \wedge \b) + (\a_2 \wedge \b),$$ and $$ \a \wedge \b = (-1)^{k-1}\b \wedge \a.$$

 We set $\Lambda^kV = 0$ for all $k > n.$  
 
 If $k \le n$  a basis for $ \Lambda^kV$ consists of $k$-vectors of the form
$$\s^{1_1} \wedge \s^{i_2} \wedge \cdots \wedge \s^{i_k}$$ where $i_1 < i_2 < \cdots i_k.$
Then  $dim \Lambda^kV  =
\left(\begin{array}{c}n \\k \end{array}\right)$

\subsection*{The $k$-vector of a polyhedral chain} (\cite{whitney}, III) A $k$-cell $\s$ determines a unique $k$-direction.  This, together with its $k$-volume $M(\s)$ determine a unique simple $k$-vector denoted $Vec(\s)$ with the same mass and direction. Define the $k$-vector of an algebraic $k$-chain $A = \sum a_i \s_i$ by $Vec(A) := \sum a_i Vec(\s_i).$    For $k = 0$ define $Vec(\sum a_i p_i): = \sum a_i.$   This definition extends to a polyhedral $k$-chain $P$ since the $k$-vector of any chain equivalent to a $k$-cell is the same as the $k$-vector of the $k$-cell.    The main purpose of introducing $Vec(A)$ in this paper is to show that the important $k$-elements defined below in Chapter \ref{discrete} are, in fact, well defined.

\begin{proposition}\label{zero}
If $P$ is a polyhedral $k$-chain then $Vec(\p P) = 0.$
\end{proposition}

\begin{proof}
This follows since $Vec(\p \s) = 0$ for every $k$-cell $\s$.
\end{proof}

\begin{theorem}\label{vec}  $Vec$ is a   linear operator
$$Vec: \cal{P}_k \to \L^k(\R^n)$$ with $$M(Vec(P)) \le M(P)$$ for all $P \in \cal{P}_k.$
\end{theorem}

\begin{proof}
This follows since $M(Vec(\s)) = M(\s)$ for every $k$-cell $\s$.
\end{proof}

\subsection*{Exterior product of $j$- and $k$-vectors}
Define $$\wedge: (\Lambda^j V) \times (\Lambda^k V) \to \Lambda^{j+k}V$$ by $$\wedge (\alpha_1 \wedge \cdots \wedge \alpha_j, \beta_1 \wedge \cdots \wedge \beta_k) = \alpha_1 \wedge \cdots \wedge \alpha_j \wedge \beta_1 \wedge \cdots \wedge \beta_k.$$  This product is distributive, associative and anticommutative:  $$\mu \wedge \lambda = (-1)^{jk} \lambda \wedge \mu.$$
 
\medskip \noindent{\bf Exercise}:   Prove that all $k$-vectors in $\Lambda^k(\R^n)$ are simple $k$-vectors for $n \le 3.$  Show this fails for $n=4$ by showing that $(e^1 \wedge e^2) + (e^3 \wedge e^4)$ is not simple.  

\subsection*{Oriented $k$-direction of a simple $k$-vector}  
\quad  \quad \quad
Define the {\itshape \bfseries $k$-direction} of a simple $k$-vector $\a$ to be the $k$-dimensional subspace of the vectors $\a^1, \cdots, \a^k$ that determine $\a = \a^1 \wedge \a^2 \wedge \cdots \wedge \a^k.$   This subspace inherits an orientation from $\a$ as follows:   Two  bases of a vector space $V$ are   equivalent if and only if the matrix relating them has positive determinant.  An equivalence class of bases is called  an {\itshape \bfseries  orientation}.  There are only two orientations of $V$ and we choose one of them and fix it, once and for all. We say a basis is {\itshape \bfseries  positively} or {\itshape \bfseries  negatively oriented}, according as to whether it has the chosen orientation.   A subspace of $V$ has no preferred orientation but it makes sense to give the $k$-direction of a simple $k$-vector the same or opposite orientation as $\a$.   If we give it the same orientation as $\a$, we obtain the {\itshape \bfseries oriented $k$-direction of $\a$}.    

\medskip

Standard operators on $V$ such as linear transformations, inner products and determinants naturally extend to corresponding structures and operators on $\Lambda^k V.$   For example, if $T:V \to W$ is linear then the transformation $\Lambda^k(T): \Lambda^k(V) \to \Lambda^k(W)$ defined on simple $k$-vectors $\a = \a_1 \wedge \a_2 \wedge \cdots \wedge \a_k$ by $$\Lambda^kT(\a) =  T(\a_1) \wedge T(\a_2) \wedge \cdots \wedge T(\a_k)$$ is linear.  Extend $\Lambda^kT$ to $k$-vectors by linearity. 
 
 \section*{Inner product spaces}  An inner product on a vector space $V$ is a real-valued function on $V \times V$ which is
 symmetric
positive definite bilinear form.  
     An example is the {\itshape \bfseries Euclidean inner product} on $\R^n$ given by $<\a,\b> = \sum a_i b_i$ where $\a = (a_1, \cdots a_n)$ and $\b = (b_1, \cdots b_n).$      An {\itshape \bfseries orthonormal basis} of $V$ consists of a basis $\s^1, \cdots, \s^n$ such that $$<\s^i, \s^j> =   \d^{ij}.$$   
 Suppose $V$ is an inner product space.  We define an inner product for simple vectors   $\a, \b \in \Lambda^kV$ as follows:
$$<\a,\b> := det(<\a_i, \b_j>)$$ where $\a = \a_1 \wedge \a_2 \wedge \cdots \a_k$ and $\b = \b_1 \wedge \b_2 \wedge \cdots \b_k.$   Since determinant is an alternating, $k$-multilinear function of the $\a'$s and $\b'$s we obtain a scalar valued function, linear in each variable
$$<\quad, \quad>: \Lambda^k V \times \Lambda^k V \to \R.$$ Note that $<\a,\b> = <\b,\a>$ since the determinant of the transpose of a matrix is the same as the determinant of the matrix.  

It is left to show that the inner product is well defined.  
    
\subsection*{Orthonormal basis of $\L^k V$}  Choose an orthonormal basis $\{e^1, \cdots e^n\}$ of $V$.  For $H = \{h_1, h_2, \cdots, h_k\}$, let $$e^H :=  e^{h_1} \wedge e^{h_2} \wedge \cdots \wedge e^{h_k} .$$ Then the collection 
of simple $k$-vectors  
$$\{e^H: H = \{h_1 < h_2 < \cdots < h_k\}\}$$ forms a basis of $\L^kV.$

\subsection*{Mass of $\a$} For a simple $k$-vector $\a$, define $M(\a) = \sqrt{<\a,\a>}.$

\subsection*{ Volume elements}   If $V$ is an inner product space, the volume element $vol$ is well defined (with respect to the chosen orientation).  It is the unique $n$-vector in $\L^nV$  with $$vol = e^1 \wedge e^2 \wedge \cdots \wedge e^n$$ where $(e^i)$ is a positively oriented orthonormal basis. Observe that $M(vol) = 1.$ 
 
 \medskip \noindent{\bf Exercise}  $\{e^H\}$ is an orthonormal basis of $\L^kV.$  (Hint:  $H \ne L \implies $the matrix has a row of zeroes.  $H = L \implies$ all but the diagonal elements vanish.   Thus $<e^H, e^J> = \pm 1$ if $H$ and $J$ contain the same indices with no repeats, and is zero, otherwise.  In particular, $vol = e^1 \wedge \cdots \wedge e^n$ is an orthonormal basis of $\L^nV$ and $<vol, vol> = 1$.   
 
 The $k$-vector $\{P_k\}$ of each $P_k$ in exercise of Chapter \ref{chainlets} is the volume element in $\R^2.$

 \section*{Two basic examples}
  
\medskip \noindent  Example 1.   The oriented $k$-direction of an oriented $k$-dimensional parellelopiped $P$ is the oriented $k$-subspace  generated by the vectors that determine $P$.  We create equivalence classes of oriented $k$-parallelopipeds $P$ in $\R^n$.  We say that  $P \sim Q$ if $P$ and $Q$ have the same  $k$-volume and oriented $k$-direction.  We take formal sums of these classes to form a vector space ${\bf P_k}$.  Clearly, there is a  1-1 correspondence between simple $k$-vectors and equivalence classes of parallelopipeds with a given oriented $k$-direction and  $k$-volume, or mass.       The notion of taking products of $k$-vectors and making $k$-parallelograms justifies the use of the term {bfserires itshape exterior product } since the $k$-parallelogram takes up space exterior to its defining vectors.  Later, we will see an interior product where the parallelogram created is interior to its defining vectors.   The term exterior, associated to the first definition given in these  notes, the exterior product, is used throughout the subject.  We will see exterior derivatives, exterior algebra, exterior chainlets.  Each time you see the term you should expect to see the exterior product playing a major role.  For example the exterior algebra is called an algebra because it comes with a product, namely the exterior product.   
\begin{theorem}
There is a canonical isomorphism   $$\Theta:\L^k(\R^n) \to {\bf P_k}$$ such that the $k$-dimensional volumes and oriented $k$-directions of $P$ and  $\Theta(P)$ are the same.  
\end{theorem}  

We assume all $k$-vectors and $k$-directions are oriented and usually omit the term ``oriented''.

This result gives us a simple geometric way to view $\L^k(\R^n)$ and will help us visualize operators and products as we go along.  However, an important feature missing from this representation is that there is no canonical representative of a $k$-vector within a class of parallelopipeds.  In particular, one cannot define the boundary of a $k$-vector.   An important goal of these notes is to resolve this problem.  

 \medskip \noindent   Example 2. The {\itshape \bfseries differential $dx^i$} of $\R^n$  is defined to be the linear functional on $\R^n$ given by $dx^i(x^1e^1  +  \cdots +  x^ne^n) := x^i.$  
  Let $D^n$  denote the linear space generated by differentials $\{dx^1, dx^2, \cdots, dx^n\}$ of $\R^n.$    The space $\L^k(D^n)$ is of special interest.  We follow the common practice of omitting the wedge sign for $k$-vectors e.g., $dx^i \wedge dx^j = dx^i dx^j.$   These $k$-vectors of differentials exist since they do so for any vector space.  The theorem tells us what they are --  linear functionals of $k$-vectors of $\R^n$.

 \begin{theorem}  The mapping $a^H dx^H \mapsto a^H e^H$ induces an isomorphism   $$\L^k(D^n) \cong  (\Lambda^k \R^n)^{\prime}.$$
 \end{theorem}
 
 \medskip

\section*{Differential forms}
A {\itshape \bfseries differential $k$-form} $\o$ on $U \subset \R^n$ is an element of the dual space of  $\Lambda^k(\R^n)$ for each $p \in U,$ $$\o:U \times \L^k(\R^n) \to \R.$$    \symbolfootnote[1]{In GC we define operators on differential forms as dual to geometric operators on $k$-vectors $\a$ defined at each point, without integrating, whereas   GMT defines operators on domains (currents) as dual to operators defined analytically on differential forms.}  In these notes, we define three basic operators on forms as dual to corresponding operators on $k$-vectors:   Hodge star $*\o$, pullback $f^* \o$  and exterior derivative $d \o:$
\begin{itemize}
\item $\star \o(p,\a) := \o(p, \star \a)$
\item $f^* \o(p,\a) := \o(p, f_* \a)$
\item $d \o(p,\a) := \o(p, \p \a)$
\end{itemize}
We are reduced to defining three geometric operators on simple $k$-vectors $\a \in \Lambda^k(\R^n)$ -- the star operator $\star \a$, pushforward $f_* \a$ and boundary $\p \a.$ Star of a simple $k$-vector will be presented in this section.
We postpone defining the boundary $\p \a$ of a simple $k$-vector $\a$ until the lecture on discrete calculus.    The idea of the boundary of a $k$-vector is a new and important concept introduced in these notes.   
\begin{theorem}\label{boundary} If $T: \R^n \to \R^n$ is a linear transformation and $\a \in \L^k(\R^n)$ then 
  $$\L^{k-1}T(\p \a) = \p( \L^kT(\a)).$$
\end{theorem}
The definitions and proof will be postponed until Lecture III as the natural norm is needed to define the boundary of a $k$-vector in $L^k(\R^n).$

\subsection*{Pullback of a differential form}
Suppose $f:U \subset \R^n \to V \subset \R^m$ is smooth.  Denote the Jacobian matrix of partial derivatives evaluated at $p \in U$ by $Df_p$. This, of course is a linear transformation $Df_p:\R^n \to \R^m.$    We will denote 
$$D^kf_p := \L^k(Df_p): \L^k(\R^n) \to \L^k(\R^n).$$  Observe that if $\a$ is a $k$-vector then $D^kf_p(\a)$ is a  well defined $k$-vector called the {\itshape \bfseries pushforward} of $\a$ under $f$.
 The {\itshape \bfseries pullback} of a $k$-form $\o$ in $V$ is a $k$-form $f^* \o$ in $U$ defined by $$f^* \o(p,\a) :=\o(f(p),  D^kf_p(\a)).$$  
 
\begin{lemma}
If $f:U \subset \R^n \to V \subset \R^m$ is smooth and $\o$ is a differential $k$-form then
\begin{enumerate}
\item $f^*(\o + \eta) = f^* \omega + f^* \eta;$
\item $f^*(\a \wedge \b) = f^* \a \wedge f^* \b;$
\item $df^* = f^* d$
\item $(g \circ f)^* = f^* \circ g^*$
\end{enumerate}
\end{lemma}
\begin{proof}\quad\\
\vspace{-.2in}
\begin{enumerate}
\item follows since $D^kf_p$ is linear.
\item follows since $D^kf_p$ acts on a $k$-vector $\a = \a^1 \wedge \dots \wedge \a^k$ by way of the linear transformation $Df_p$ applied to each $\a^i$.
\item $$df^*\o(p,\a) = f^*\o(p, \p(\a)) = \o(f(p), D^{k-1}f_p(\p \a)).$$ On the other hand, by Theorem \ref{boundary} 
$$\begin{aligned} f^*d \o(p,\a) = d \o (f(p), D^kf_p(\a)) &=   \o(f(p), \p D^kf_p( \a)) \\&= \o(f(p), D^kf_p(\p \a)).\end{aligned}$$   
\item $$
\begin{aligned} (g \circ f)^*\o(p,\a) &= \o(g\circ f(p), D^k (g\circ f)_p(\a)) \\&= 
\o(g\circ f(p), D^k g_{f(p)} D^kf_p(\a)) \\&= f^* \circ g^* \o(p,\a).
\end{aligned}$$
\end{enumerate}   
\end{proof}

Examples.  
1.  Let $f:\R^2 \to \R$ be defined by $f(x,y) = x-y$  and $\o = dt$ a 1-form on $\R.$   Then 
$$\begin{aligned}
f^*\o((x,y), (a^1, a^2)) &= \o (f(x,y), Df_{x,y}(a^1,a^2)) \\&= dt(x-y, a^1 -a^2) \\&= a^1-a^2 = (dx -dy)(a^1,a^2).
\end{aligned}$$  Therefore $f^* dt = dx -dy.$  

Remark.  The standard approach (commonly unjustified) is 
$$t = x-y \implies dt = dx - dy.$$

2. Let $f(t): \R^1 \to \R^2$ be defined by   $f(t) = (t^2, t^3).$  Let $\o$ be the $1$-form in $\R^2$ given by $\o = x dy$.   Then 
$$\begin{aligned}
f^*\o(t, 1) &= \o(f(t), Df_t(1)) \\&= \o((t^2,t^3), (2t, 3t^2)) \\&= xdy((t^2,t^3), (2t, 3t^2)) \\&= t^2 3t^2 \\&= 3t^4dt(t,1).
\end{aligned}
$$  Therefore $f^*(xdy) = 3t^4 dt.$

\subsection*{Hodge star dual}

Given  a simple $k$-vector $\a$, we wish to find a canonical simple $(n-k)$-vector $\star\a$ such that $\a \wedge \star\a = M(\a)^2 vol.$ 
 
We further require that the $(n-k)$-direction of $\star\a$ be perpendicular to the $k$-direction of $\a$.    
We are reduced to showing existence and uniqueness.

For this, we recall the  Riesz  representation theorem:
\begin{theorem}\label{rep}
Suppose $V$ is an inner product space and $f \in V^*$ is a linear functional.  There exists a unique $w \in V$ such that $$f(v) = <v,w>.$$  
\end{theorem}
\begin{proof}
Let $e^1, \cdots, e^n$ be orthonormal.  Set $b_i:= f(e^i).$  Set
$w = \sum  b_j e^j.$ Then $<e^i,w> = b_i = f(e^i).$
\end{proof}

Now fix $\a \in \L^kV$.  For $\b \in \L^{n-k}$ the mapping $\b \mapsto \a \wedge \b$ is a linear transformation $$\L^{n-k} \to \L^n.$$  The latter space has dimension one.  Therefore one may define $ f_{\a}(\b)$ by  $$\a \wedge \b = f_{\a}(\b) vol$$ where $f_{\a}$ is a linear functional on $\L^{n-k}V.$  According to Theorem \ref{rep} there exists a unique $(n-k)$-vector $\star\a$ such that $$\a \wedge \mu = <\star\a,\mu>vol$$ for all $\mu \in \L^{n-k}.$  
\begin{lemma} For simple $\a$ 
the $(n-k)$-direction of $\star\a$ is the orthogonal complement of the $k$-direction of $\a$, oriented so that $\a \wedge \star\a$ is positively oriented.  
\end{lemma}

In particular 
\[
\begin{aligned} \a \wedge \star\a &=(-1)^{k(n-k)}\star\a \wedge \a = (-1)^{k(n-k)} <\star\star\a,\a>vol \\&=(-1)^{k(n-k)} (-1)^{k(n-k)}<\a,\a>vol = M(\a)^2 vol.
\end{aligned}
\]

\begin{lemma} If $\a, \b$ are $k$-vectors then
\begin{enumerate}
\item $\star\star\a = (-1)^{k(n-k) }\a.$
\item $\a \wedge \star\b = \b \wedge \star\a =  <\a,\b>vol$

\end{enumerate}

\end{lemma}

 If $\o$ is a differential $k$-form, define the Hodge star dual $\star\o$ as $$\star\o(p,\a):= \o(p, \star\a)$$ for all $(n-k)$-vectors $\a.$  If $\o$ is a $k$-form we know that $\o = \o^1 \wedge \cdots \wedge \o^k$ where the $\o^i$ are $1$-forms.   Each $\o^i$ corresponds to a vector $w^i$ from the above exercise.  
\begin{proposition}  Assume $\o = \o^1 \wedge \cdots \wedge \o^k$ is a simple $k$-form. Then  
$(\star\o)(v^1 \wedge \cdots \wedge v^{n-k})$ is the volume of the $n$-parallelogram spanned by  $v^1, \cdots, v^{n-k}, w^1, \cdots w^k.$
\end{proposition}

\chapter{Differential forms and chainlets}\label{forms} 
\renewcommand{\thetheorem}{3.\arabic{theorem}}
\renewcommand{\thesection}{3.\arabic{section}}
\setcounter{theorem}{0}
\setcounter{section}{0}
\section*{Isomorphisms of differential forms and cochains}
We recall two classical results from integral calculus:
\begin{theorem}[Classical Stokes' theorem] \label{classicStokes} If $P$ is a polyhedral  $k$-chain and $\o$ is a smooth $k$-form defined in a neighborhood of $P$ then
$$\int_{\p P} \o = \int_P d\o.$$
\end{theorem}  

\begin{theorem}[Classical change of variables]\label{classicchange} If $P$ is a polyhedral  $k$-chain, $\o$ is a smooth $k$-form and $f$ is an orientation preserving diffeomorphism  defined in a neighborhood of $P$ then
$$\int_{f P} \o = \int_P f^* \o.$$

\end{theorem}

  \subsection*{The flat norm}
 
 Whitney's flat norm \cite{whitney} on polyhedral chains $A \in \cal{P}_k$ is defined as follows:
$$|A|^{\flat} = \inf\{M(B) + M(C): A = B + \p C,  B \in \cal{P}_k, C \in \cal{P}_{k+1}\}.$$

{\itshape \bfseries Flat $k$-forms}  (\cite{whitney}, 12.4) are characterized as all bounded measurable $k$-forms $\o$ such that there exists a constant $C > 0$ such that $\sup |\int_{\s} \o | < CM(\s)$ for all $k$-cells $\s$ and  $\sup |\int_{\p \t} \o | < CM(\t)$ for all $(k+1)$-cells $\t$.     The exterior derivative $d\o$ of a flat form $\o$ is defined a.e. and satisfies $$\int_{\p \t} \o = \int_{\t} d\o.$$

\medskip \noindent{\bf Exercises} 

1. Prove that the form  
$$\o = \begin{cases} dx -dy  , &x >y\\ 0, &x \le y
\end{cases}$$   is flat but the form 
$$\eta = \begin{cases} dx + dy, &x >y\\ 0 &x \le y
\end{cases}$$ is not flat.   (Hint.  Consider small squares with centers on the diagonal.)  Thus $\o$ is smoothly homotopic to a form that is not flat.   Show that there are forms arbtirarily close to $\o$ that are not flat.  Show that the components of $\o$ are not flat.

We conclude that the flat topology is limited at the level of chains and cochains. (Comment for experts:  Some problems disappear at the level of homology and cohomology.)   

2.  Let $P$ be the boundary of a rectangle with bottom edge length 10 and side length 0.1.  Let $Q$ be the same curve without one of the short edges.  Show that $|P|^{\flat} < |Q|^{\flat}.$

The {\itshape \bfseries support} of a differential form is the closure of the set of all points $p \in \R^n$ such that $\o(p)$ is nonzero.   Let $U$ be an open subset of $\R^n$   and $\o$ be a bounded measurable $k$-form whose support is contained in    $U$.  In what follows, let $\s$ denote a $k$-cell and $\t$ a $(k+1)$-cell.
  
    Define $$\|\o\|_0 := \sup \left\{\frac{\int_{\s}\o}{M(\s)}: \s \subset spt \o \right\}.$$ 
  Inductively define $$\|\o\|_r := \sup\left\{\frac{\|\o -T_v\o\|_{r-1}}{|v|}:  spt(\o -T_v \o) \subset U\right\}.$$ 
  Define $$\|\o\|_0^{\prime} := \sup \left\{\frac{\int_{\p \t}\o}{M(\t)}: \t \subset spt \o\right\}$$  and   $$\|\o\|_r^{\prime} := \sup\left\{\frac{\|\o -T_v\o\|_{r-1}^{\prime}}{|v|}:  spt(\o -T_v \o) \subset U\right\}.$$

Define $$|\o|_0 := \|\o\|_0$$ and  for $r \ge 1$,
  $$|\o|_r := \max\{\|\o\|_o, \cdots, \|\o\|_r, \|\o\|_0^{\prime}, \cdots, \|\o\|_{r-1}^{\prime}\}.$$
   We say that $\o$ is of class $B^r$ if $|\o|_r < \i.$  Let $ \cal{B}_k^r$ denote the space of differential $k$-forms of class $B^r.$

 \begin{lemma}
 If $|\o|_1 < \i$ then $d \omega$ is defined a.e.  Furthermore,  \[\int_{\p \s} \omega = \int_{\s} d\omega\] 
\end{lemma}

\begin{proof}
 If $|\o|_1 < \i$ then $\omega$ is a flat form.  It follows from    (\cite{whitney}, 12.4) that $d\omega$ is defined a.e. and satisfies Stokes' theorem on cells. 
\end{proof} 

  \begin{lemma}\label{omega} If $\o \in \cal{B}_k^r$, $r \ge 1$, then $$|\o|_r = \max\{\|\o\|_o, \cdots, \|\o\|_r, \|d\o\|_0 , \cdots, \|d\o\|_{r-1} \}.$$  
\end{lemma}

Therefore 
 
\begin{equation}\label{equation} |d\o|_{r-1} \le |\o|_r.
\end{equation}
\subsection*{Remark} On a smooth manifold these norms are equivalent to the $C^{r-1 +Lip}$ norms of analysis.  That is, the $(r-1)$-derivatives exist and satisfy Lipschitz conditions.  
However, the norms can also be defined on Lipschitz manifolds where one cannot speak of higher derivatives.  We only know that derivatives of Lipschitz functions exist almost everywhere.  However,   a constant function  is of class $B^r$ for all $r$, even on  a Lipschitz manifold.  This opens the possibility of extending chainlet geometry to Lipschitz Riemannian manifolds which we will develop in a later chapter.  (We must be able to define divergence free vector fields, exterior derivative and star operator a.e.)   

The next result  generalizes the standard integral inequality of calculus:
$$\left|\int_P \o\right| \le M(P)|\o|_o$$ where $P$ is polyhedral and $\o$ is a bounded, measurable form.

\begin{theorem}[Fundamental integral inequality of chainlet geometry]\label{oldintegral} Let $P \in \cal{P}_k$, $r \in \Z^+,$ and $\o \in \cal{B}_k^r$ be defined in a neighborhood of $spt(P).$ Then
$$\left|\int_P \o\right| \le |P|^{\natural_r}|\o|_r.$$
\end{theorem}
 
\begin{proof}    We first prove $ \left|\int_{\s^{j}} \o \right| \le \|\s^j\|_{j}\|\o\|_{j}.$
Since $\|\o\|_0 = |\o|_0$ we know
$$\left|\int_{\s^0} \o \right| \le M(\s^0)|\o|_0 = \|\s^0\|_0\|\o\|_0. $$    
 
 Use the change of variables formula  \ref{classicchange} for the translation $T_{v_j}$ and induction   to deduce
$$\begin{array}{rll} \left|\int_{\s^{j}} \o \right| =  \left|\int_{\s^{j-1} - T_{v_j}\s^{j-1}} \o \right| 
&=  \left|\int_{\s^{j-1}} \o - T_{v_j}^* \o \right|  \\&\le  \|\s^{j-1}\|_{j-1}\|\o - T_{v_j}^*\o\|_{j-1} \\&\le  \|\s^{j-1}\|_{j-1}\|\o\|_{j}|v_{j}| \\&= \|\s^j\|_{j}\|\o\|_{j}\end{array}$$
 
  By linearity $$\left|\int_{D^j} \o\right| \le \|D^j\|_j \|\o\|_j$$
for all   $D^j \in \cal{D}_k^j$.

  We again use induction to prove $\left|\int_P \o\right| \le  |P|^{\natural_r}|\o|_r.$ 
We know   $\left|\int_P \o\right| \le |P|^{\natural_0}|\o|_0.$  
Assume the estimate holds for $r-1.$ 

  Let $\e > 0$. There exists $P = \sum_{j=0}^r  D^j  + \p C$ such that $|P|^{\natural_r} >
\sum_{j=0}^r \|D^j\|_j + |C|^{\natural_{r-1}} - \e$. By Stokes' theorem for polyhedral chains, inequality (\ref{equation}) and
induction
 $$\begin{array}{rll}  \left|\int_P \o\right| &\le \sum_{j=0}^r \left|\int_{D^j} \o \right| + | \int_C d\o| \\& \le \sum_{j=0}^r \|D^j\|_j\|\o\|_j + |C|^{\natural_{r-1}} |d\o|_{r-1}\\& \le (\sum_{j=0}^r
\|D^j\|_j  + |C|^{\natural_{r-1}}) |\o|_r\\&\le  (|P|^{\natural_r} +
\e) |\o|_r.

\end{array} $$
Since the inequality holds for all $\e > 0$ the result follows.
\end{proof} 
 
\begin{corollary} $ |P|^{\natural_r}$ is a norm on the space of polyhedral chains $\cal{P}_k$. 
\end{corollary}

\begin{proof} Suppose $P \ne 0$ is a polyhedral chain. There exists a smooth differential form $\o$   such that $\int_P  \o \ne 0$. Then
$0 <  \left|\int_P \o \right| \le |P|^{\natural_r}|\o|_r$ implies $|P|^{\natural_r} > 0.$ 
\end{proof}

The Banach space of polyhedral    $k$-chains $\cal{P}_k$ completed with the
norm $|\quad |^{\natural_r}$ is denoted $\cal{N}^r_k $. The elements of $\cal{N}^r_k $ are called
{\itshape \bfseries  $k$-chainlets of class} $N^r$.

It follows from Proposition \ref{lem.boundary} that the boundary $\p A$ of a k-chainlet $A$ of class $N^r$ is well defined as a $(k - 1)$-chainlet of class $N^{r+1}$.   If $P_i \to A$  in the $r$-natural norm define $$\p A:= \lim_{i \to \i} \p P_i.$$   By Theorem \ref{oldintegral} the integral $\int_A \o$ is well defined for k-chainlets $A$ of class $N^r$ and differential $k$-forms of class $B^r$.  If $P_i \to A$  in the $r$-natural norm define $$\int_A \o:= \lim_{i \to \i} \int_{P_i} \o.$$ 
\begin{theorem}\label{stokes}[Generalized Stokes' theorem] If $A$ is a $k$-chainlet of class $N^r$ and $\o$ is a $(k-1)$-form of class $B^{r+1}$ defined in a neighborhood of $spt A$ then 
$$\int_{\p A} \o = \int_A d\o.$$
\end{theorem}

\begin{proof}  Choose polyhedral $P_i \to A$ in the $r$-natural norm.  By Stokes' theorem for polyhedral chains
$$\int_A d\o = \lim \int_{P_i} d\o = \lim \int_{\p P_i} \o = \lim \int_{\p A} \o.$$
\end{proof}

Later, in the discrete chapter, we give a proof without relying on the classical Stokes' theorem for polyhedral chains.  

\medskip \noindent{\bf Examples of chainlets}
\begin{enumerate}
\item {\itshape \bfseries The boundary of any bounded, open subset $U$ of $\R^n$.}
One may easily verify that the boundary of any bounded, open set $U \subset \R^n$, such as the Van Koch snowflake, supports a well defined chainlet $B= \p U$ of class $N^1$.  Suppose the frontier  of $U$ has positive Lebesgue area.     Then the chainlet $B' = \p(\R^n - \overline{U})$ has the same support as $B$, namely the frontier of $U$, but since $B+B'$  bounds a chainlet with positive mass it follows that $B$ and $-B'$ are distinct. 
      
 \item  {\itshape \bfseries Graphs of functions} The graph  of a nonnegative $L^1$ function $f:K \subset \R^n \to \R$ supports a chainlet $\G_f$ if $K$ is compact.  This can be seen by approximating $\G_f$ by the polyhedral chains $P_k$ determined by a sequence of step functions $g_k$ approximating $f$.  The difference $P_k - P_{k+j}$ is a $1$-difference $k$-cellular chain.   The {\itshape \bfseries subgraph} of a nonnegative function $f$ is the area between the graph of $f$ and its domain.  Since the subgraph of $f$ has finite area, it follows that $\|P_k - P_{k+j}\|_1 \to 0$ as $j, k \to \infty$.  Hence, the sequence $P_k$ is Cauchy in the $1$-natural norm.   The boundary $\p \G_f$ is a chainlet that identifies the discontinuity points of $f$.   
\end{enumerate}

Later, we will define the natural norms and develop a full theory of integration from first principles, without reference to the classical Stokes' theorem or even the Riemann integral. 

\section*{ The Banach space ${\cal N}_k^{\i}$ of chainlets}
It can be easily shown that the $r$-natural norms satisfy the
inequalities
$$|P|^{\natural_0}  \ge  |P|^{\natural_1}  \ge
|P|^{\natural_2}
\ge
\cdots$$  for any polyhedron  $P$.    It follows that there are natural inclusion     mappings $\eta_r: {\cal
N}_{k}^r \to {\cal
N}_{k}^{r+1}$  defined as follows:  If $A \in {\cal
N}_{k}^r$ then $A = \lim P_j$ in the $r$-natural norm.  Define $\eta_r(A) := \lim P_j$ in the $(r+1)$-natural norm.   
  
 For $P$ a polyhedral chain define $$|P|^{\natural_{}} := \lim_{r \to \i} |P|^{\natural_r}.$$ This limit exists   since the $r$-natural norms are decreasing.   
  
 \begin{theorem}   $|\quad|^{\natural_{}} $ is a norm on polyhedral chains.
 \end{theorem} 
 
 \begin{proof}    
 If $P,Q \in \cal{N}_k^{\i}$ then  $$
 \begin{array}{rll} |P +Q|^{\natural_{}} &= \lim_{r \to \i} |P+Q|^{\natural_r} \\&\le   \limsup_{r \to \i} (|P|^{\natural_r} + |Q|^{\natural_r}) \\&=  \lim_{r \to \i} |P|^{\natural_r} +  \lim_{r \to \i} |Q|^{\natural_r} \\&= |P|^{\natural_{}}+|Q|^{\natural_{}}
 \end{array}
 $$ and
 $$|\l P|^{\natural_{ }}=  \lim_{r \to \i} |\l P|^{\natural_r} = |\l|  \lim_{r \to \i} |P|^{\natural_r} = |\l||P|^{\natural_{}}.$$
Clearly, $P = 0 \implies |P|^{\natural_{}} = 0.$   It remains to show $|P|^{\natural_{}} \ne 0 \implies P \ne 0 $ for a polyhedron $P$.  But if $P$ is a nonzero polyhedral chain, there exists a smooth form $\o$ and a constant $C > 0$ such that $\int_P \o \ne 0$ and $|\o|_r < C$  for all $r$. Then $0 < \left|\int_P \o\right| \le |P|^{\natural_r}|\o|_r < C |P|^{\natural_r}$ for each $r$.   It follows that $|P|^{\natural} \ge  \left|\int_P \o\right|/C > 0$.   
\end{proof}
We call elements of the Banach space $\cal{N}_k^{\i}$ obtained upon completion {\itshape \bfseries  $k$-chainlets of class $N^{\i}$} or, more simply, {\itshape \bfseries $k$-chainlets}.

\begin{theorem}  If $A$ is a chainlet in $\R^n$ and $v \in \R^n$ then
$$|A -T_vA|^{\natural_{}} \le |v||A|^{\natural_{}}.$$
\end{theorem}

Hence translation of a chainlet converges  uniformly to 0 as a function of $A$ and $v$.  

\begin{proof}
Let $P_i \to A$.  Then $P_i - T_v P_i \to A - T_v A.$  It follows that 
$$|A -T_vA|^{\natural_{}}  = \lim |P_i -T_vP_i|^{\natural_{}}  \le |v||P_i|^{\natural_{}} \to |v||A|^{\natural_{}}.$$
\end{proof}
 
\begin{theorem}
$$|\p A|^{\natural_{}} \le |A|^{\natural_{}}.$$
\end{theorem}

\begin{proof}
$$|\p A|^{\natural_{}}  = \lim_{r \to \i}|\p A|^{\natural_{r}} \le 
\lim_{r \to \i}| A|^{\natural_{r-1}} = |A|^{\natural_{}}.$$
\end{proof}
 
In these lectures we introduce a number of operators on chainlets all of which are bounded in the chainlet spaces.  The Banach spaces of chainlets are not reflexive since they are separable and differential forms are not separable \cite{continuity}.  Thus $\cal{N}^{\natural_{}}$  is strictly contained in $(\cal{N}^{\natural_{}})^{**}$.  Now operators dual to operators on differential forms automatically send chainlets into the double dual space.   It is surprising that the operators map $\cal{N}^{\natural_{}}$ into itself.

\subsection*{Characterization of the Banach space of chainlets}
\begin{theorem}
Suppose $|\quad|^{\prime}$ is a seminorm of polyhedral chains satisfying 
\begin{enumerate}
 \item $|\p D^0|^{\prime} \le |D^0|^{\natural_{r-1}}$ and
\item $|D^i|^{\prime} \le \|D^i\|_i$  for all multicellular chains $D^i$ and $0 \le i \le r$. 
\end{enumerate}
   Then $|P|^{\prime} \le |P|^{\natural_{r}}$ for all polyhedral chains $P$.
\end{theorem}

\begin{proof}
 
For $r \ge 1$, if $\e > 0$  there exists a decomposition $P = \sum_{i=0}^{r} D^i + \p C$ such that 
\[
\begin{aligned}
|P|^{\natural_{r}} >  \sum_{i=0}^{r} \|D^i\|_i +  |C|^{\nat_{r-1}} -\e.
\end{aligned}
\]
By assumption $$|P|' \le  \sum_{i=0}^{r} |D^i|' + |\p C|' \le  \sum_{i=0}^{r} \|D^i\|_i + |C|^{\natural_{r-1}} < 
|P|^{\natural_{r}} + \e.$$  The result follows 
\end{proof}
 
  \begin{corollary}\label{natnorms}  $$|A|^{\nat_r} = \inf\left\{\sum_{i=0}^r \|D^i\|_i + |C|^{\nat_{r-1}}: A = \sum_{i=0}^r D^i + \p C, D^i  \in \cal{N}^{\nat_r}, C  \in \cal{N}^{\nat_{r-1}}\right \}.$$  
\end{corollary}
\begin{proof}
Denote the rhs by $|A|_{<r>}$.  This is clearly a seminorm.   By the characterization of the $r$-natural norm, we know that $  |A|_{<r>} \le |A|^{\nat_r} .$  

Let $\e > 0$.  There exists $A = \sum D^i + \p C,  D^i  \in \cal{N}^{\nat_r}, C \in   \cal{N}^{\nat_{r-1}}$ such that $$|A|_{<r>} > 
\sum_{i=0}^r \|D^i\|_i + |C|^{\nat_{r-1}} - \e.$$  
Then $$\begin{aligned} |A|^{\nat_r} &\le \sum|D^i|^{\nat_r} + |C|^{\nat_{r-1}} 
\\&=  \sum|D^i|_r + |C|^{\nat_{r-1}} \\&\le \sum |D^i|_i   +|C|^{\nat_{r-1}} 
\\&\le \sum \|D^i\|_i + |C|^{\nat_{r-1}}  \\&< |A|_{<r>} + \e. \end{aligned}$$

\end{proof}

 \begin{corollary}
 
The natural norm is the largest seminorm in the class of seminorms satisfying 
\begin{enumerate}
\item $|A|^{\prime} \le M(A)$,
 \item $|\p A|^{\prime} \le |A|^{\prime}$ and
\item $|A-T_vA|' \le |v||A|'$.
\end{enumerate}
 
\end{corollary}

\begin{proof}  First observe that conditions (1) and (3) imply 
$|D^i|^{\prime} \le \|D^i\|_i$  for all multicellular chains $D^i$ and $0 \le i \le r$. 
We know $|D^0|^{\prime}  \le \|D^0\|_0$ by (1).   Assume the claim holds for $i-1.$  By (3) and the induction hypothesis
$$ |\s^i|^{\prime}  = |\s^{i-1} -T_v \s^{i-1}|^{\prime} \le   |v||\s^{i-1}|^{\prime} \le  |v|\|\s^{i-1}\|_{i-1} = \|\s^{i}\|_i .$$ The claim is established by taking linear combinations.

 Let $P$ be a polyhedral chain.   Given $\e > 0$ there exists $r$ such that $$|P|^{\natural_{}} > |P|^{\natural_{r}} -\e/2$$  and a decomposition $P = \sum D^i + \p C $ such that $$|P|^{\natural_{r}} > \sum \|D^i\|_i + |C|^{\natural_{r-1}} - \e/2.$$  Hence 
$$|P|^{\natural_{}} > \sum \|D^i\|_i + |C|^{\natural_{r-1}} - \e.$$  If $k = n$ then the same inequality holds, except we set $C = 0$ since $C$ has dimension $n+1.$   In this case we have $$|P|' \le \sum |D^i|' \le \sum \|D^i\|_i < |P|^{\natural_{}} + \e.$$  Since the result holds for all $\e > 0$ we conclude $|P|' \le |P|^{\natural_{}}.$
 Now assume the result holds for polyhedral $k$-chains $C$, we prove it holds for polyhedral $(k-1)$-chains $P$.  

 Thus by induction
\[
\begin{aligned} |P|' &\le \sum |D^i|^{\prime} + |\p C|^{\prime} \\&\le  \sum \|D^i\|_i + | C|^{\prime} \\&\le 
\sum \|D^i\|_i + |C|^{\natural} \\&\le  \sum \|D^i\|_i + |C|^{\natural_{r-1}} \\&< |P|^{\natural_{}} +\e.
\end{aligned}
\]
\end{proof}

Let $X$ be a Banach space with norm $|\quad|^{\prime}$.  We say an operator $S: X  \to X$ is {\itshape \bfseries is Lipschitz bounded }   if  there exists $K > 0$ such that $|S(A)|^{\prime} \le  K|A|^{\prime}$ for all $A \in X$.
An operator $S: X \times \R^n \to X$ is {\itshape \bfseries Lipschitz bounded}  if  there exists $K > 0$ such that $|S(A,v)|^{\prime} \le  K|v||A|^{\prime}$ for all $A \in X$ and $v \in R^n.$   Two examples that interest us are the translation operator $S(A,v) = A -T_vA$ and the boundary operator $S(A) = \p A.$   
We assume $K = 1$ in what follows as different constants lead to the same Banach space.

Let $X^0$ be the completion of the space of polyhedral chains with the mass norm.  \begin{corollary}
The Banach space of chainlets is the smallest Banach space containing $X^0$ and which has Lipschitz bounded boundary and  translation operators.
\end{corollary}
  
\subsection*{Characterization of cochains as differential forms}

The $r$-natural norm of a cochain $X \in ({\cal N}^{r})^{\prime}$ is defined by
$$|X|^{\natural_r} := \sup_{P \in {\cal P}}\frac{|X \cdot P|}{|P|^{\natural_r}}.$$
The differential operator $d$  on cochains is defined as the dual to the
boundary operator $dX \cdot A := X \cdot \p A.$     It remains to show how cochains relate to integration of differential forms and how the operator d given above relates to the standard exterior derivative of differential forms.   
  If
$X \in ({\cal N}_k^{r})^{\prime}$ then $dX \in ({\cal N}_{k+1}^{r-1})^{\prime}$
by Lemma \ref{lem.boundary}.

\subsection*{Cochains and differential forms}  In this section we show the operator $\Psi$ mapping differential forms of class $B^r$ into the dual space of chainlets of class $N^r$ via
integration
$$\Psi(\o)
\cdot A :=
\int_A \o$$ is a norm preserving isomorphism of graded algebras.      
 
It follows from   Theorem \ref{oldintegral} that$\Psi(\o) \in (\cal{N}_k^r)^{\prime}$ with
                            $$|\Psi(\o)|^{\natural_r} \le |\o|_{r}.$$

\begin{theorem}[Extension of the theorem of de Rham] \label{theorem.iso}     Let $r \ge 0.$ To each cochain $X \in
\left({\cal N}_k^{r}\right)^{\prime}$ there corresponds a
unique differential form
$\phi(X)
\in {\cal B}_k^{r} $ such that $\int_{\s}\phi(X) = X\cdot \s$  for all
cells $\s$.
This correspondence is an isomorphism with
$$ |X|^{\natural_r} = |\phi(X)|_{r}.$$
 
 If $r \ge 1$ then
   $$\phi(dX) = d\phi(X).$$
 \end{theorem}

This is proved in \cite{iso}.   

\begin{corollary} \label{theorem.char} If $A, B \in {\cal N}_k^{r}$
satisfy $$\int_A \o =
\int_B
\o$$ for all $\o \in {\cal B}_k^{r}$ then $A = B$.
\end{corollary}

\begin{proof}  Let $X \in  ({\cal N}_k^{r})^{\prime}. $  By Theorem \ref{theorem.iso} the form $\phi(X)$ is of class $B^r$.  Hence
 $$X \cdot (A-B) = \int_{A-B}\phi(X) = 0.$$ It
follows that $A=B.$
\end{proof}

\begin{corollary} \label{theorem.norm} If $A \in   \cal{N}_k^r$ then

$$|A|^{\natural_r} = \sup\left\{\int_A \o :   \o \in B^r_k, |\o|_{r} \le 1\right\}.$$

\end{corollary}

\begin{proof}  By Theorem \ref{theorem.iso}

\[
\begin{array}{rll}
|A|^{\natural_r} &= \sup\left\{\frac{|X \cdot A|}{|X|^{\natural_r}}: X \in  (\cal{N}_k^r)^{\prime}\right\} \\&= 
\sup\left\{\frac{|\int_A \phi(X)|}{|\phi(X)|_{r}}: \phi(X)
\in   {\cal B}_k^{r}\right\} \\&=   \sup\left\{\frac{|\int_A
\o|}{|\o|_{r}}: \o\in   {\cal
B}_k^{r}\right\}.
\end{array}
\]
\end{proof}

\subsection*{Cup product} Given a $k$-cochain $X$ and a $j$-cochain $Y$, we
define their {\itshape \bfseries cup product} as the $(j+k)$-cochain  
$$X\cup Y := \Psi(\phi(X) \wedge \phi(Y)).$$
The next result follows directly from Theorem
\ref{theorem.iso}.

\begin{lemma}  Given  $X \in  (\cal{N}_k^r)^{\prime}$ and $Y \in  (\cal{N}_j^r)^{\prime}$  the cochain $X\cup Y \in  ({\cal
N}_{k+j}^{r}(R^n))^{\prime}$  with
                            $$|X\cup Y|^{\natural_r} = |\phi(X) \wedge \phi(Y)|_{r}.$$
Furthermore
                                $$\phi(X \cup Y) = \phi(X) \wedge \phi(Y).$$

\end{lemma}

\begin{theorem} If
  $X \in
({\cal N}_k^{r})^{\prime}, Y \in
({\cal N}_j^{r})^{\prime}, Z \in
({\cal N}_{\ell}^{r})^{\prime},$  and  $f
\in \cal{B}_0^{r+1}$ then
\begin{itemize}
\item[(i)] $|X \cup Y|^{\natural_r}
\le |X|^{\natural_r}|Y|^{\natural_r}$;
 \item[(ii)]    $d(X \cup Y) = dX \cup Y + (-1)^{j+k }  X \cup
dY;$
\item[(iii)]  $(X \cup Y) +
(Z \cup  Y) = (X+Z)\cup Y;$ and
\item[(iv)] $a(X \cup Y) = (aX \cup Y) = (X \cup aY).$

\end{itemize}
\end{theorem}

\begin{proof} These follow by using the isomorphism of
differential forms and cochains Theorem \ref{theorem.iso} and then applying corresponding results for
differential forms and their wedge products.

\end{proof}

Therefore the isomorphism $\Psi$ of Theorem \ref{theorem.iso} is one on graded
algebras.

 \subsection*{Continuity of {$ \mathbf {Vec(P)}$}}  

\begin{lemma}\label{Riemann}
Suppose $P$ is a polyhedral chain   and $\o$ is a bounded, measurable differential form.  If $\o(p) = \o_0$ for a fixed covector $\o_0$ and for all $p,$ then $$\int_P \o = \o_0 \cdot Vec(P).$$
\end{lemma} 

\begin{proof}
This follows from the definition of the Riemann integral.
\end{proof}

 \begin{theorem}\label{massr}
If $P$ is a polyhedral $k$-chain and  $r \ge 1$ then  \[M(Vec(P)) \le |P|^{\nat_r}. \]
If $spt(P) \subset B_{\e}(p)$  for some $p \in \R^n$ and  $ \e > 0$ then 
\[|P|^{\nat_1} \le M(Vec(P)) + \e M(P) .\]    
\end{theorem}

\begin{proof}
Set $\alpha = Vec(P)$  and let $\eta_0$ be a covector such that $|\eta_0|_0 = 1$, and $\eta_0 \cdot \alpha = M(\alpha)$. Define the $k$-form $\eta$ by $\eta(p, \beta)  := \eta_0(\beta) .$  Since $\eta$ is constant it follows that $\|\eta\|_r = 0$ for all $r > 0$  and $\|d\eta\|_r = 0$ for all $r \ge 0.$  Hence $|\eta|_r = |\eta|_0  = |\eta_0|_0= 1.$
  By Lemma \ref{Riemann} and Theorem \ref{oldintegral} it follows that 
\[M(Vec(P)) = \eta_0 \cdot Vec(P)  = \int_P \eta \le    |\eta|_r |P|^{\natural_{r}} = |P|^{\natural_{r}}.\]

For the second inequality we use Corollary \ref{theorem.norm}.  It suffices to show that $\frac{|\int_P \o|}{|\o|_1}$    is less than or equal the right hand side for any $1$-form $\omega$ of class $B^1$.   Given such $\omega$ define the $k$-form        $\omega_0(q, \beta)  := \o(p,\beta) $ for all $q$. By Lemma \ref{Riemann} 
\[
\begin{aligned}
 \left|\int_P \o\right| &\le \left|\int_P \omega_0\right| + \left|\int_P \omega -\omega_0\right |  \\&\le
 |\o(p) \cdot Vec(P)| + \sup_{q\in spt P}|\o(p) -\o(q)| M(P) \\& \le
   \|\o\|_0M( Vec(P))  +  \e \|\o\|_1 M(P) \\& \le
   |\o|_1( M(Vec(P)) + \e M(P))    \end{aligned}
\]
\end{proof}

If $A = \lim_{i \to \infty} P_i$ in the $r$ natural norm then $\{P_i\}$ forms a Cauchy sequence in the $r$-natural norm.   By Theorem \ref{massr} $\{Vec(P_i)\}$ forms a Cauchy sequence in the mass norm on $\L^k(\R^n).$   Define \[Vec(A) := lim Vec(P_i).\]  This is independent of the choice of approximating $P_i$, again by Theorem \ref{massr}.
\begin{corollary} \label{massrcor}  \[Vec: \cal{N}_k^r \to \L^k(\R^n)\] is linear and continuous. 
\end{corollary}
   
\begin{corollary} \label{massrcor2}
Suppose $A$ is a chainlet of class $N^r$ and $\omega$ is a differential form of class $B^r.$ If $\o(p) = \omega_0$ for a fixed covector $\omega_0$ and for all $p$  then \[\int_A \omega = \omega_0 \cdot Vec(A).\] 
\end{corollary}

\begin{proof}  This is merely Lemma \ref{Riemann} if $A$ is a polyhedral chain.
   Theorem \ref{massr} lets us take limits in the $r$-natural norm.  If $P_i \to A$ in $\cal{N}_k^r$ then by Corollary \ref{massrcor} $Vec(P_i) \to Vec(A).$  Therefore
   \[ \int_A \o = \lim_{i \to \infty} \int_{P_i} \o = \lim_{i \to \infty} \o_0 \cdot Vec(P_i) = \o_0 \cdot Vec(A).\]
\end{proof}

 \subsection*{The supports of a  cochain and of a chainlet}   The
support $spt( X)$ of a cochain $X$ is the set of points $p$ such that for each $\e > 0$ there is a cell $\s  \subset
U_{\e}(p)$ such that $X\cdot \s \ne 0.$

 The support $spt(A)$ of a chainlet $A$ of class $N^r$  is the set of points $p$ such that for each $\e > 0$ there is a cochain $X$ of class $N^r$ such that $X \cdot A \ne 0$ and $X \cdot \sigma = 0 $ for each $\sigma$  supported outside $U_{\e}(p).$   We prove that this coincides with the definition of  the support  of $A$ if $A$ is a polyhedral chain. Assume $A = \sum_{i=1}^m a_i \sigma_i$ is nonoverlapping and the $a_i$ are nonzero.  We must show that $spt(A)$ is the union $F$ of the $spt(\sigma_i)$ using this new definition.    Since $X \cdot A = \int_{A}  \phi(X)$  it follows that $spt(A) \subset F.$   Now suppose $x \in F$; say $x \in \sigma_i.$  Let $\e > 0.$  We find easily a smooth differential form $\omega$ supported in $U_{\e}(p)$, $\int_{\sigma_i} \omega \ne 0$, $\int_{\sigma_j} \omega = 0, j \ne i$.  Let $X$ be the cochain determined by $\omega$ via integration.   Then $X \cdot A \ne 0$     and $X \cdot \sigma = 0 $ for each $\sigma$  supported outside $U_{\e}(p).!
$  

\begin{proposition}\label{spt}  If $A$ is a chainlet  of class $N^r$ with  $spt(A)= \emptyset$ then $A = 0.$  If $X$ is a cochain of class $N^r$ with   $spt(X) = \emptyset $ then $ X = 0.$
\end{proposition}

\begin{proof}  By Corollary \ref{theorem.norm} suffices to show $X \cdot A = 0$ for any cochain $X$ of class $N^r$.  Each $p \in spt(X)$ is in some neighborhood $U(p)$ such that $Y \cdot A = 0$ for any $Y$ of class $N^r$ with $\phi(Y) = 0$ outside $U(p)$.    Choose a locally finite covering  $\{U_i, i \ge 1\}$  of $spt(X)$.  Using a partition of unity $\{\eta_i\} $  subordinate to this covering we have $$X = \sum \eta_i X$$ and $\phi(\eta_i X) = \eta_i \phi(X) = 0$ outside $U_i$.  Hence $$X \cdot A = \sum (\eta_iX \cdot A) = 0.$$

For the second part it suffices to show that $X \cdot \s = 0$ for all simplexes $\s$.  Each $p \in \s$ is in some neighborhood $U(p)$ such that $X \cdot \t = 0$ for all $\t \subset U(p).$  We may find a subdivision $\sum \s_i$ of $\s$ such that each $\s_i$ is in some $U(p)$.  Therefore $X \cdot \s = \sum X \cdot \s_i = 0.$
\end{proof}  

\subsection*{Cochainlets}

 The elements of the dual space $(\cal{N}_k^{\i})^{\prime}$ to $\cal{N}_k^{\i}$  are called {\itshape \bfseries cochainlets}.   
All the results on cochains of class $N_k^r$ carry over to cochainlets by taking limits. Cochainlets are characterized as differential forms, each  with a uniform bound on all the derivatives of its coefficient functions. 
 These form a Banach space and a differential graded module, but not an algebra.  For example $e^x$ is a function in $(\cal{N}_0^{\i})^{\prime}$, but $e^x \cdot e^x = e^{2x}$ is not since its  derivatives are not uniformly bounded.   We contrast this with $C^{\i}$ forms which are a differential graded algebra but not a Banach space.  
 On the other hand, each $(\cal{N}_k^{r})^{\prime}, 0 \le r < \i,$ is both a Banach space and a differential graded algebra.  
 
\chapter{Calculus on fractals -- star operator}\label{fractals}
\renewcommand{\thetheorem}{4.\arabic{theorem}}
\renewcommand{\thesection}{4.\arabic{section}}
\setcounter{theorem}{0}
\setcounter{section}{0}
 
\subsection*{$k$-elements} In this section we make precise the notion of an {\itshape \bfseries infinitesimal} of calculus.   Imagine taking an infinitely thin square card and cutting it into four pieces.  Stack the pieces and repeat, taking a limit.  What mathematical object do we obtain?  The reader will recall Dirac monopoles  which are closely related.  We show the limit, which the author calls a {\itshape \bfseries    $k$-element}, exists as  a well defined chainlet and thus may be acted upon by any chainlet operator.      We emphasize that these operators have geometric definitions, as opposed to the weak definitions arising from duals of differential forms.
Let $p \in \R^n$  and $\alpha$ be a    $k$-direction in $\R^n$.    A  {\itshape \bfseries unit    $k$-element} $\alpha_p$ is defined as follows:
For each $\ell \ge 0$, let $Q_{\ell} = Q_{\ell}(p,\alpha)$ be the weighted $k$-cube centered at $p$ with $k$-direction $\alpha$, edge $2^{-\ell}$ and  coefficient $2^{k \ell}$.  
Then
$M(Q_{\ell}) =1$ and $Vec(Q_{\ell}) = \a.$ 
  We show that $\{Q_{\ell}\}$ forms a Cauchy sequence in the $1$-natural norm.    Let $j \ge 1$ and estimate $|Q_{\ell} -Q_{\ell + j}|^{\natural_1}.$  Subdivide

 $Q_{\ell}$ into
$2^{kj}$ binary cubes $Q_{\ell,i}$ and consider $Q_{\ell +j}$ as $2^{kj}$ copies of   $\frac{1}{2^{kj}}Q_{\ell +j}.$   We form  $1$-difference $k$-cells    of these subcubes of
$Q_{\ell} -Q_{\ell + j}$ with translation distance $\le 2^{-\ell}.$   Since the  mass of each
$Q_{\ell}$ is one, it follows that 
\[| Q_{\ell} -Q_{\ell + j}|^{\natural_1}  = \left|
\sum_{i = 1}^{2^{kj}} \left(Q_{\ell,i} - \frac{1}{2^{kj}}Q_{\ell + j}\right)\right|^{\natural_1} \le \sum_{i = 1}^{2^{kj}} \|Q_{\ell,i} - \frac{1}{2^{kj}}Q_{\ell + j}\|_1
\le 
  2^{-\ell}.\]
  
 Thus $Q_{\ell}$ converges to a $1$-natural chain denoted $\alpha_p$  with $|\alpha_p - Q_{\ell}|^{\natural_1} \le 2^{1-\ell }.$   If we let $\alpha$ be any simple $k$-vector with nonzero mass, the same process will produce a chainlet $\alpha_p$ depending only on $\alpha$ and $p$, whose mass is the same as that of $\alpha$ and supported in $p$.   We obtain

\begin{equation}\label{eq2} \alpha_p = \lim Q_{\ell} \mbox{ and } |\alpha_p - Q_{\ell}|^{\natural_1} \le 2^{1-\ell }M(\a).
\end{equation}

  Since $Vec(Q_{\ell}) = \alpha$ for all $\ell$, it follows from Corollary \ref{massrcor} that $Vec(\a_p) = \a.$   If $\omega$ is a form of class $\cal{B}^1$ defined in a neighborhood of $p$ then $\int_{\alpha_p} \omega = \o(p; \alpha) $ by   Corollary \ref{massrcor2}.

 \begin{proposition}\label{vecprop}  For each  nonzero simple  $k$-vector $\alpha$ and $p \in \R^n$  there exists a unique chainlet $\alpha_p \in \cal{N}_k^1$ such that   $Vec(\alpha_p) = \alpha,$ $spt(\alpha_p) = \{p\}$ and $\int_{\alpha_p} \o = \o(p; \alpha)$ for all forms $\o$ of class $\cal{B}_k^1.$
\end{proposition}

\begin{proof} Let $\a_p = \lim Q_{\ell}$ be as in (\ref{eq2}).   It is unique by  Corollary \ref{theorem.char} since 
 $\int_{\alpha_p} \o = \o(p; \alpha)$ for all forms $\o$ of class $\cal{B}_k^1.$   Since  $Vec(\alpha_p) = \alpha$ we know $\alpha_p \ne 0.$  Since $spt Q_{\ell} \subset B_p(2^{-\ell})$ then 
  $spt \alpha_p$ is either the empty set or the set $\{p\}.$  By Proposition \ref{spt} $spt A = \emptyset \implies A = 0.$    Hence $spt A = \{p\}.$   
\end{proof}
The next proposition tells us that the particular shapes of the approximating polyhedral chains to $\a_p$ does not matter.  There is nothing special about cubes.  
\begin{proposition}
Let $\{P_i\}$ be a sequence of polyhedral $k$-chains such that \[M(P_i) \le C,   spt(P_i) \subset B_{\e_i}(p), Vec(P_i) \to \alpha\] for some $C > 0$ and $\e_i \to 0$.  Then $P_i \to \a_p$ in the $1$-natural norm. 
\end{proposition}
 
\begin{proof}  By (\ref{eq2})  $\a_p = \lim Q_{i}$ with $Vec(Q_i) = \a$.  By Theorem \ref{massr}  and Corollary \ref{massrcor}
\[
\begin{aligned} |P_i - Q_i |^{\natural_{1}} &\le M(Vec(P_i) -  Vec(Q_i)) + \e_i M(P_i - Q_i)
   \to 0.
\end{aligned}
\]
\end{proof}

\begin{theorem} Fix $p\in \R^n.$ The operator \[Vec: \cal{N}_k^r \to \L^k(\R^n)\] is one-one on chainlets supported in   $p$.
 \end{theorem}

\begin{proof}    By Proposition \ref{vecprop} and 
Theorem \ref{vec}   we only need to show that if $A \in \cal{N}_k^r$ which is supported in $p$ and satisfies $Vec(A) = 0$ then $A = 0.$  Let $X$ be an $r$-natural cochain.  Define $X_0$ by \[\phi(X_0)(q) := \phi(X)(p)  \mbox{ for all } q.\]     By Corollary \ref{massrcor2} 
\[X \cdot A = X_0 \cdot A = \phi(X)(p) \cdot Vec(A) = 0\] implying $A = 0.$ 
\end{proof}

 A {\itshape \bfseries $k$-element chain} $\dot{P} = \sum_{i=1}^m b_i (\a_p)_i$   is a chain of   $k$-elements $ (\a_p)_i$ with coefficients $b_i$ in $G$. (Note that both the $k$-vector $\a$ and point $p$ may vary with $i$.)   Denote the vector space of $k$-element chains in
$\R^n$ by $\cal{E}_k.$    The next theorem is a quantization of chainlets including, for example, fractals, soap films, light cones and manifolds.

\begin{theorem}[Density of element chains] \label{thmdiscrete} The space of  $k$-element chains $\cal{E}_k$ is dense in $\cal{N}_k^r.$
\end{theorem}

\begin{proof}   Let $R$ be a unit $k$-cube in $\R^n$ centered at $p$ with $k$-direction $\a.$   For each $j \ge 1 $  subdivide $R$
into $2^{kj}$ binary cubes  $R_{j,i}$ with midpoint $p_{j,i}$  and edge $2^{-j}.$  Since $R_{j,i} = 2^{-jk} Q_j(p_{j,i},\a)$  it follows that

 \[
\begin{array}{rll} |R_{j,i} -   2^{-jk}  \alpha_{p_{j,i}}|^{\natural_1} &\le   2^{-jk}|Q_j(p_{j,i},\alpha) -    \alpha_{p_{j,i}}|^{\natural_1} \\&\le  
 2^{-jk} 2^{-j+1} =  2^{-j+1}M(R_{j,i}).\end{array}\]  Let $\dot{P}_j = \sum_{i=1}^m 2^{-jk}\alpha_{p_{j,i}}.$    Then 
\[|R -  \dot{P}_j|^{\natural_1} \le    2^{-j+1}\sum M(R_{j,i}) = 2^{-j+1}M(R) = 2^{-j+1}.\]

This demonstrates  that  $\dot{P}_j  \buildrel \natural_{1} \over \to R$.  This readily extends to any cube with edge $\e$. 

Use the Whitney decomposition to subdivide a $k$-cell $\tau$ into binary $k$-cubes.   For each $j \ge 1$ consider the finite sum of these cubes    with edge $\ge 2^{-j}.$    Subdivide each of these cubes into subcubes $Q_{ji}$ with edge $2^{-j}$  obtaining  $\sum_i Q_{ji} \to \tau $ in the mass norm as $j \to \i$.      Let $\alpha = Vec(\tau)$ and $p_{ji}$ the midpoint of $Q_{ji}.$    Then 
\[|\tau - \sum_i \alpha_{p_{ji}}|^{\natural_{1}} \le |\tau - \sum_i  Q_{ji}|^{\natural_{1}}  +
\sum_i| Q_{ji} -  \alpha_{p_{ji}}|^{\natural_{1}}.\]    We have seen that the first term of the right hand side tends to zero as $j \to \i.$  By (\ref{eq2}) the second is bounded by $\sum_i M(Q_{ji}) 2^{-j+1} \le M(\tau) 2^{-j+1} \to 0.$    It follows that 
$\tau$ is approximated by elementary $k$-chains   in the $1$-natural norm. 
   Thus elementary $k$-chains are dense in $\cal{P}_k.$  The result follows since polyhedral chains
are dense in chainlets.
\end{proof}

{\subsection*{Multiplication of a smooth chainlet by a function}  
 Let $\a_p $ be a $k$-element and $f$ a smooth function defined in a neighborhood of $p$.   Define $$f \a_p :=  f(p) \a_p.$$  Extend to element chains $\dot{P} = \sum a_i (\a_p)_i$ by $$f \dot{P} := \sum a_i f (\a_p)_i.$$   It follows immediately from the definitions that $\int_{f\a_p} \o = \int_{\a_p}f \o.$  Therefore $$\int_{f \dot{P}} \o = \int_{\dot{P}} f \o.$$
\begin{theorem}  Let $\dot{P} $ be a $k$-element chain and $f$ a smooth function defined in a neighborhood of $\dot{P}$.   Then 
$$|f \dot{P}|^{\natural_{r}} \le 2^r|f|_r | \dot{P}|^{\natural_{r}}.$$
\end{theorem}

\begin{proof}  By the chain rule and the Fundamental Integral Inequality of chainlet geometry, 
 $$ \left|\int_{f \dot{P}} \o\right|   = \left|\int_{\dot{P}}f \o\right| \le |\dot{P}|^{\natural_{r}}|f \o|_r \le   |\dot{P}|^{\natural_{r}}2^r |f|_r|\o|_r.$$     By \ref{theorem.norm} 
 $$|f \dot{P}|^{\natural_{r}} = \sup\frac{|\int_{f \dot{P}} \o|}{|\o|_r} \le 2^r |f|_r |\dot{P}|^{\natural_{r}}.$$
 \end{proof}

If  $A = \lim P_i$ is a chainlet and $f$ a function  define $$fA:= \lim f P_i.$$

\subsection*{Geometric star operator} Recall the Hodge star operator $\star$ of differential forms $\o$.    We next define a geometric star operator on chainlets.     If $\alpha$ is a simple $k$-vector in $\R^n$ then $\star \alpha$ is defined to be the simple $(n-k)$-vector with $(n-k)$-direction orthogonal to the $k$-direction of $\alpha$, with complementary 
orientation and with $M(\alpha) = M(\star \alpha)$.    The operator $\star$ extends to   $k$-element chains $\dot{P}$ by linearity.   It follows immediately that $\star \o(P; \star \alpha) = \o(p;  \alpha).$  (Indeed, we prefer to define $\star \o$ in this way, as dual to the geometric $\star.$)  Hence $\int_{ \dot{P}} \omega =
\int_{\star \dot{P}} \star \omega$.  According to Theorem \ref{theorem.norm}
$|\star \dot{P}|^{\natural_r} = |\dot{P}|^{\natural_r}.$    We may therefore define $\star A$ for any chainlet $A$ of class $N^r$ as follows:  By Theorem \ref{thmdiscrete} there exists $k$-element chains $\{\dot{P}_j\}$ such that $A = \lim_{j \to \i} \dot{P}_j$ in the $r$-natural norm.   Since  $\{\dot{P}_j\}$ forms a Cauchy sequence we know $\{\star \dot{P}_j\}$ also forms a Cauchy sequence.  Its limit in the $r$-natural norm is denoted $\star A.$    This definition is independent of the choice of the sequence $\{\dot{P}_j\}.$

 \begin{theorem}[Star theorem]   \label{thm.star}
$\star  : {\cal N}_k^{r} \to{\cal N}_{n-k}^{r} $ is a
 norm-preserving linear  
operator that is adjoint to the Hodge star operator on forms.  It  satisfies $\star  \star  =  (-1)^{k(n-k)} I$ and 
 
 $$\int_{\star  A}
\o = (-1)^{k(n-k)} \int_A \star \o$$   for all $A \in \cal{N}_k^r$ and
 all $(n-k)$-forms $\o$ of class $B^r, r \ge 1,$ defined in a neighborhood of $spt(A).$
\end{theorem}
This result was first announced in \cite{madeira} and will appear in \cite{hodge}
   
  \begin{proof} We first prove this for   $k$-elements $\alpha_p.$   Since  $\alpha_p$ is a $1$-natural chainlet we may integrate $\omega$ over it.  Hence \[\int_{\alpha_p} \omega = \o(p;\alpha) = \star \o(p; \star \alpha) = \int_{\star \alpha_p} \star \o.\] It follows by linearity that $ \int_{\dot{P}}  \omega = \int_{\star \dot{P}} \star \omega $ for any elementary $k$-chain $\dot{P}$.  Let $A$ be a chainlet of class $N^r$. It follows from Theorem \ref{thmdiscrete} that $A$ is approximated by elementary $k$-chains $ A = \lim_{j \to \i} \dot{P}_j$ in the $r$-natural norm.  We may apply continuity of the integral (Theorem \ref{oldintegral}) to deduce 
   \[\int_{A} \omega = \int_{\star A} \star \o.\]

The Hodge star operator on   forms satisfies $\star \star  \o = (-1)^{k(n-k)} \o.$  It follows that $$\int_{A} \star \o  = \int_{\star A} \star \star  \o = (-1)^{k(n-k)}\int_{\star A} \o.$$

   \end{proof}

\subsection*{Geometric coboundary of a chainlet}
Define the {\itshape \bfseries geometric  coboundary } operator $$\diamondsuit: \cal{N}_k^r\to \cal{N}_{k+1}^{r+1}$$  by $$\diamondsuit  : =
\star  \p
\star.$$ Since $\p^2 = 0$  and $\star \star = \pm I$ it follows that $\diamondsuit^2 = 0.$
 
The following theorem follows immediately from properties of boundary $\p$ and star $\star $.  Let $\d:=  (-1)^{nk +n +1}\star d \star$ denote the coboundary  operator on differential forms.  

 \begin{theorem}[Coboundary operator theorem] $\diamondsuit : \cal{N}_k^r \to {\cal N}_{k+1}^{r+1}$ is a nilpotent linear operator
satisfying
 \begin{itemize}
\item[(i)] $\int_{\diamondsuit A} \omega = (-1)^{n+1} \int_A  \delta \omega$  for all $\omega$ defined in a neighborhood of $spt(A)$;
\item[(ii)] $\star  \p  = (-1)^{n+k^2 +1}\diamondsuit\star;$ and
\item[(iii)] $|\diamondsuit A|^{\natural_r} \le |A|^{\natural_{r-1}}$ for all chainlets $A$. 
\end{itemize} 
 \end{theorem}
 
\subsubsection*{Geometric interpretation of the coboundary of a chainlet} This has a geometric
interpretation seen by taking  approximations by polyhedral chains.   For example,  the
coboundary
of $0$-chain $Q_0$ in $\R^2$ with  unit $0$-mass and supported in
a single point $\{p\}$  is the limit of $1$-chains $P_k.$

The coboundary of a $1$-dimensional unit cell $Q_1$   in $\R^3$  is
approximated by a ``paddle wheel'',
 supported in a neighborhood of  $|\s|$.

If  $Q_2$ is a unit $2$-dimensional square in
$\R^3$ then its coboundary $\diamondsuit Q_2$ is approximated by the sum of
two weighted sums of oppositely oriented pairs of small $3$-dimensional
balls, 
one collection slightly above
$Q_2$, like a mist, the other collection slightly below $Q_2.$     A snake approaching the
boundary of a lake knows when it has arrived.  A bird
approaching the coboundary of a lake knows when it has
arrived.
 
\subsection*{Geometric Laplace operator}

The geometric Laplace operator \[\Delta: {\cal N}_k^{r}
\to {\cal N}_k^{r+2}\] is defined on chainlets
by \[\Delta  := (\p + \diamondsuit)^2  = (\p \diamondsuit + \diamondsuit \p).\]

    Let $\square$ denote the Laplace operator on  differential forms. 

\begin{theorem}[Laplace operator theorem] Suppose $A \in {\cal N}_k^{r}$ and  $\o \in {\cal
B}_k^{r+2} $ is defined in a neighborhood of $spt(A).$ Then $\Delta A \in  {\cal N}_k^{r+2}$,  $$|\Delta A|^{\natural{r+2}} \le |A|^{\natural_r},$$ and $$\int_{\Delta A} \o = (-1)^{n-1}\int_{A} \square \o.$$ 
\end{theorem}

The geometric Laplace operator on chainlets requires at least the
$2$-natural norm.   Multiple iterations of
$\Delta$ require   the $r$-natural norm for larger and larger $r$.
For spectral analysis and applications to dynamical systems the normed linear space
${\cal N}_{k}^{{\i}}$  with the operator  $$\Delta: {\cal N}_{k}^{{\i}} \to {\cal
N}_{k}^{{\i}}$$ should prove useful.  

A chainlet is   {\itshape \bfseries harmonic} if   $$\Delta A = 0.$$  It should be of considerable interest to 
study the spectrum of the   geometric Laplace operator $\Delta$ on chainlets.\symbolfootnote[1]{The
geometric Laplace operator was originally defined by the author with the object of developing a geometric Hodge theory.}

 \subsection*{Geometric representation of
differentiation of distributions}
 
  An {\itshape \bfseries $r$-distribution} on $\R^1$ is a bounded
linear functional on functions $f \in {\cal B}_0^{r}(\R^1)$  with compact support. 
Given a one-dimensional  chainlet $A$ of class $N^r$ 
define the  $r$-distribution
$\th_A$  by  $\th_A(f) := \int_A f (x)dx$,   for
$f \in {\cal B}_0^{r}(\R^1).$

\begin{theorem}  $\theta_A$ is linear and injective.  Differentiation in the sense of distributions
corresponds geometrically to the operator $ \star \p $.   That is,  $$\th_{\star \p A} = (\th_A)^{\prime}.$$\symbolfootnote[2]{Since this paper was first submitted in 1999, the author has extended this result to currents. \cite{currents}} 
\end{theorem}

\begin{proof}
Suppose $\th_A = \th_B$.  Then $ \int_A
f(x)dx = \int_B f(x)dx$ for all functions $f \in
{\cal B}_0^{r}.$  But all $1$-forms $\omega \in {\cal
B}_1^{r}$ can be written $\omega = f dx.$  By Corollary \ref{theorem.char}
chainlets are determined by their integrals and    thus  
   $A = B$.
 
  We next show that $\th_{\star \p A} =
(\th_A)^{\prime}.$  Note that $\star (f(x) dx) = f(x).$  Thus
\[
\begin{array}{rll}
\th_{\star \p A}(f) &= \int_{\star \p A} f(x)dx  = \int_{\p A}
f = \int_A df \\&= \int_A f^{\prime}(x)dx =
\th_A(f^{\prime}) = (\th_A)^{\prime}(f).
\end{array}
\]
\end{proof}

\section*{Extensions of   theorems of Green and Gauss}

\subsection*{Curl of a vector field over a chainlet}

Let $S$ denote a smooth, oriented surface with boundary in
$\R^3$ and $F$ a smooth vector field defined in a neighborhood of $S$.  The usual way to integrate the curl of a vector
field $F$ over $S$ is to integrate the Euclidean dot product of
curl$F$ with the unit normal vector field of $S$ obtaining    $\int_S curlF \cdot n dA$.  By the curl theorem this integral equals
$\int_{\partial S} F \cdot  d\s.$ 

We translate this into the language of chainlets and differential forms.

Let $\omega$ be the unique differential $1$-form associated to $F$
by way of the Euclidean dot product.   The differential form version of $curl F $ is    $\star d\o.$  
The  unit normal vector field of $S$ 
 can be represented as the chainlet $\star S$.  Thus the net curl of $F$ over
$S$ takes the form  $\int_{\star S} \star  d\o.$ By the Star theorem \ref{thm.star} and Stokes' theorem for chainlets \ref{stokes}  this integral equals $\int_S d\omega = \int_{\p S}
\o.$ The vector version of the right hand integral is  $\int_{\partial S} F \cdot  ds. $  The following extension of Green's curl theorem to chainlets of arbitrary dimension and codimension follows immediately from Stokes' theorem and the Star theorem and is probably optimal.

\begin{theorem}[Generalized Green's curl theorem]  Let $A$ be a  $k$-chainlet of class $N^r$ and $\omega$ a differential $(k-1)$-form of class $B^r$ defined in a neighborhood of $spt(A).$  Then \[\int_{\star A} \star  d \omega =
\int_{\p A}
\o.\] 
\end{theorem}  
 
\begin{proof}
This is a direct consequence of  Theorems \ref{stokes} and \ref{thm.star}.
\end{proof}
 
It is not necessary for  tangent spaces  to exist for $A$ or $\p A$ for this theorem to hold.

 \subsection*{Divergence of a vector field over a chainlet} \medskip The usual way to calculate divergence of a vector field
$F$ across a boundary of a smooth surface
$D$ in $\R^2$  is to integrate the dot product of $F$ with
the unit normal vector field of
$\partial D$.   According to
Green's  Theorem, this quantity equals  the
integral of the divergence of $F$ over $D$.  That is, 
\[\int_{\partial D} F \cdot nd\sigma = \int_D divF dA.\]   Translating this into the language of differential forms and chainlets with an appropriate sign adjustment, we
  replace the unit normal vector field over $\p D$ with the chainlet $\star  \p D$ and $div F$ with the differential form $d \star \o.$ 
 We next give an extension of the Divergence theorem to $k$-chainlets in $n$-space.  As before, this follows immediately from Stokes' theorem and the Star theorem and is probably optimal.

\begin{theorem}[Generalized Gauss divergence theorem]  
Let $A$ be a  $k$-chainlet of class $N^r$   and $\omega$ a differential $(n-k+1)$-form of class $B^{r+1}$    defined in a neighborhood of $spt(A)$
then
\[\int_{\star  \partial A} \omega = (-1)^{(k-1)(n-k-1)} \int_{A} d\star \o.\]
\end{theorem}

\begin{proof}
This is a direct consequence of  Theorems \ref{stokes} and \ref{thm.star}.
\end{proof}

As before, tangent vectors need not be defined for the theorem to be valid and it holds in all dimensions and codimensions.

\subsection*{Manifolds}  The diffeomorphic image $\phi_*C$  in 
Euclidean space $\R^n$ of a $k$-cell $C$ in $\R^n$ supports a unique $k$-chainlet for which integrals of $k$-forms coincide.       A simple proof to this uses the implicit function theorem.   The image is locally the graph of a smooth function and all such graphs naturally support chainlets.  Therefore every diffeomorphism $\phi:  U\to V$ of open sets in $\R^n$ induces a linear map from 
chainlets in $U$ to chainlets in $V$, commuting with the boundary and pushforward operators.  If $W$ is a 
coordinate domain in a smooth manifold $M$, it then makes sense to speak of 
{\em chainlets in $W$}, meaning the image of a chainlet in $\R^n$ under a diffeomorphism from an 
open set in $\R^n$ to $W$, for this is independent of the diffeomorphism.  Now define a 
chainlet in $M$ to mean a finite sum of chainlets in coordinate domains.  More 
accurately, these are {\em smoothly embedded} chainlets.   Immersed chainlets can also be defined.  Thus to each smooth manifold corresponds a family of Banach spaces 
 of chainlets, and to smooth maps there 
correspond linear maps of these spaces.   

All this is done without fixing Riemannian metrics.  Stokes theorem follows by using partitions 
of unity and the theorem for $\R^n$.  With metrics, star, curl, divergence, etc. can be 
introduced.

   Chainlets   may also be defined on Lipschitz Riemannian manifolds.  
%

    In the definitions give above of the natural chainlet norms,     replace cells $\s^0$ with singular cells  $\t^0 = f\s^0$.   Define $M(\t^0)$  to be $k$-dimensional Hausdorff measure of $\t^0$.      Replace vectors $v$ with smooth, divergence free vector fields $v$ defined in a neighborhood of $spt \tau$ and let  $T_v$ denote the time one map of the flow of $v$.  Define
\[|v| := \sup\{|v(p)|: p \in spt v\}.\]   Norms of differential forms are defined as before, replacing vectors with vector fields.   The previous definitions and results carry through locally.  Global results require the pushforward operator be defined for chainlets and a change of variables result.  These are   naturally established using   $k$-elements and deduced for chainlets by taking limits.     
 
If 
$f:U\subset \R^n \to V \subset \R^n$ is Lipschitz   
then $Df_p$ is defined on a subset of full measure   by Rademacher's theorem.  Therefore, for a.e. $p \in U,$  
$$f_*(\a_p) :=  Df^k_p(\a_p)$$    is well defined so that $f^* \o(p, \a_p) = \o(f(p), f_* \a_p).$   Passing to integrals, we have $$\int_{f_* \a_p} \o = \int_{\a_p} f^*\o,$$  yielding 

\begin{proposition}\label{change} $$\int_{f_* \dot{P}} \o = \int_{\dot{P}} f^*\o$$ for all element chains $\dot{P}.$   
\end{proposition}

\begin{theorem}\label{pullback}
If $f:U \subset \R^n \to V \subset \R^n$ is a biLipschitz metric preserving mapping and $\o$ is a form of class $\cal{B}_k^r$ then $$|f^* \o|_r \le |f|_{Lip}^{k+r} |\o|_r.$$
\end{theorem}

\begin{proof}  Observe $M(f \s) \le |f|^k_{Lip} M(\s)$ for any $k$-cell $\s$.  Then 
$$ \frac{\int_{\s} f^* \o}{M(\s)} =  \frac{\int_{f\s}  \o}{M(\s)} \le \frac{M(f \s) \|\o\|_0}{M(\s)} \le |f|_{Lip}^k  \|\o\|_0.$$  Thus 
$$\|f^* \o\|_0 \le |f|_{Lip}^k \|\o\|_0.$$
Since $f$ is biLipschitz the pushforward $f_*v$ is well defined.  
Since $f$ is metric preserving if $v$ is divergence free, so is $f_* v.$
Then
$$\begin{aligned}\frac{\int_{\s} f^* \omega - T_v f^* \omega}{|v|M(\s)} &=\frac{\int_{f \s}   \omega - T_{f_*v}  \omega}{|v|M(\s)} \\&\le \frac{M(f \s) \|\omega - T_{f_* v} \omega\|_0 }{|v|M(\s)} \\&\le |f|_{Lip}^k \frac{ \|\omega - T_{f_*v} \omega\|_0 }{|v|} \\&\le |f|_{Lip}^k \|\omega\|_1\frac{|f_*v|}{|v|}.\end{aligned}$$ 

Hence $$\|f^* \omega\|_1 \le |f|_{Lip}^{k+1} \|\omega\|_1.$$

By induction $$\frac{\|f^* \omega - T_v f^* \o\|_{r-1}}{|v|} = \frac{\|f^*(\o - T_{f_*v} \o)\|_{r-1}}{|v|} \le |f|_{Lip}^{k+r-1}  \frac{\|\o - T_{f_*v} \o\|_{r-1}}{|v|}.$$ It follows that $$\|f^*\o\|_r \le |f|_{Lip}^{k+r}\|\o\|_r.$$

Finally, note that $$\|d f^* \o\|_{r-1} = \|f^* d\o\|_{r-1} \le |f|_{Lip}^{k+r} \|d \o\|_{r-1}.$$ The result follows from Lemma \ref{omega}.

\end{proof}

\begin{proposition}  Let $f: U \subset \R^n \to \R^n$ be a biLipschitz, metric preserving mapping  and $\dot{P}$ be a $k$-element chain.  Then  $$|f_*\dot{P}|^{\natural_{r}} \le |f|_{Lip}^{k+r}|\dot{P}|^{\natural_{r}}.$$  
\end{proposition}

\begin{proof}  By Theorems \ref{oldintegral} and \ref{pullback} $$\left|\int_{f_*\dot{P}} \o \right|  = \left| \int_{\dot{P}} f^* \o\right| \le |f^*\o|_r |\dot{P}|^{\natural_{r}} \le  |f|_{Lip}^{k+r}|\o|_r|\dot{P}|^{\natural_{r}}.$$

Hence \[ |f_* \dot{P}|^{\natural_{r}} = \sup \frac{\left| \int_{f_*\dot{P}} \o\right|}{|\o|_r} \le |f|_{Lip}^{k+r} |\dot{P}|^{\natural_{r}}\]
\end{proof}

Since element chains are dense in chainlets we may define $$f_*A := \lim_{i \to \i} f_* \dot{P_i}$$ where $A = \lim_{i \to \i} \dot{P_i}$ in the $r$-natural norm.   We deduce
\begin{theorem}[Pushforward operator]\label{pushforward}
Let $f: U \subset M \to M$ be a biLipschitz metric preserving mapping and $A$ be a chainlet of class $N^r$.  Then  $$|f_*A|^{\natural_{r}} \le |f|_{Lip}^{k+r}|A|^{\natural_{r}}.$$ 
\end{theorem}
We close this section with a concise and general change of variables formula.  
\begin{theorem}[Change of variables]\label{change2}   Let $f: U \subset \R^n \to \R^n$ be a biLipschitz metric preserving mapping, $\o$ be a differential form of class $B^r$ and $A$ be a chainlet of class $N^r$. Then $$\int_{f_*A} \o = \int_A f^* \o.$$
\end{theorem}

This follows by Proposition \ref{change} and taking limits in the chainlet norm.

 This permits much of chainlet geometry to be extended to Lipschitz Riemmanian manifolds.

 \subsection*{Heat equation for chainlet domains}  
Uniqueness of solutions: Let $A$ be a chainlet in $\R^n.$
Let $R = A \times [0,b].$  Then $$\p R = \p A \times [0,b] + (-1)^n A \times b + (-1)^{n-1} A \times 0.$$  
Consider the heat equation 
$$\Delta u = \sum \frac{\p ^2 u}{\p x_i^2} = \frac{\p u}{\p t}.$$  Suppose $u$ vanishes on $\p A \times [0,b]$ and on $A \times 0.$  Now $dt = 0$ on $A \times b$ since $t = b.$ 
Hence 
$$\int_{\p R} u *dudt = 0.$$   

Set $$\b = 2u(*du)dt + (-1)^{n-1} u^2 dv$$ where $dv = dx_1dx_2 \cdots dx_n.$  Since $$du \wedge (*du) = (grad \, u)^2 dv = \sum \left(\frac{\p u}{\p x_i}\right)^2 dv$$ we have
$$d \b   = 2(grad \, u)^2 dvdt.$$  Therefore, by Stokes' theorem for chainlet domains $$\int_{\p R} \b = 2\int_R  (grad \, u)^2 dvdt.$$  

Hence 
$$(-1)^{n-1}\int_{\p R} u^2 dv= 2\int_R (grad \, u)^2 dv dt.$$ I.e., 
$$\int_{A \times b}u^2 dv + 2 \int_R(grad\, u)^2dv dt = 0$$ and thus 
$$(grad \, u)^2 = \sum \left(\frac{\p u}{\p x_i}\right)^2 = 0,$$ implying  $$\frac{\p u}{\p x_i} = 0$$ on $R$. Hence $u$ is identically $0$ on $R$.    We conclude that if two temperature distributions coincide initially at $t = 0$ and always on $\p R$ then they must be the same at each point of $R$ and for each $t$.
\subsection*{Magnetic field for a chainlet domain} (draft) (e.g., the Sierpinski gasket)
  A magnetic field $B$ in absence of an electric field satisfies a Maxwell equation 
$$curl(B) = (4¹/c)j,$$ where $j$ is the current and $c$ is the speed of light. How do we get the magnetic field $B$, when the current is known? Stokes theorem can give the answer: take a closed path $C$ which bounds a chainlet surface $S$. The line integral of $B$ along $C$ is the flux of $curl(B)$ through the surface. By the Maxwell equation, this is proportional to the flux of $j$ through that surface. 

Simple case of the Biot-Savard law. Assume $j$ is contained in a wire of thickness $r$ which we align on the z-axis. To measure the magnetic field at distance $R > r$ from the wire, we take a curve $C : r(t) = (Rcos(t), Rsin(t), 0)$ which bounds a disc $S$ and measure $$2\pi RB = C B \cdot ds  = S curl(B) dS = 4 S \pi /c j dS = 4 \pi J/c,$$  where $J$ is the total current passing through the wire. The magnetic field satisfies $B = 2J/(cR)$.   \footnote{In the second part of these notes the author will present a formulation of Maxwell's equations using $k$-elements and the star and boundary operators.}  

\subsection*{Green's formula for chainlets}
 In the following theorem, the forms $\o$ and $\eta$ satisfy Sobolev conditions, $M$ is a smooth manifold with boundary.  $\bf{t}\o$ is the tangential component to $\p M;$  $\bf{n} \eta$ is the normal component.  
\begin{theorem}[Green's formula]
Let $\o \in W^{1,p} \Omega^{k-1}(M)$ and $\eta \in W^{1,q}\Omega^k(M)$ be differential forms on $M$ where $\frac{1}{p}+ \frac{1}{q} =1.$  Then
$$<d\o,\eta> = <\o, \d \eta> + \int_{\p M} \bf{t} \o \wedge \star \bf{n} \eta.$$
\end{theorem}
  In this section, we give a version of Green's formula that assumes smooth forms and chainlet domains.

\begin{theorem}[Chainlet Green's Formula]
Let $\o \in \cal{B}_{k-1}^{r+1}$ and $\eta \in \cal{B}_k^{r+1}.$ If $A \in \cal{N}^r(M)$ then
$$ \int_A d \o \wedge \star \eta - \int_A \o \wedge d \star \eta =  \int_{\p A} \o \wedge \star \eta.$$
\end{theorem}

\begin{proof}
This follows from Stokes' theorem for chainlets:
$$\int_A d(\o \wedge \star \eta) = \int_{\p A} \o \wedge \star \eta.$$  But 
$$\int_A d(\o \wedge \star \eta) = \int_A d \o \wedge \star \eta - \int_A \o \wedge d \star \eta.$$
\end{proof}

Green's formula is often stated assuming Sobolev conditions on the form.  In the second part of these  lecture notes, we will give Sobolev $W^{1,p}$ versions of the natural norms by weighting the difference norms with the constant $p$.

\chapter{Poincar\'e duality}\label{duality}
\renewcommand{\thetheorem}{5.\arabic{theorem}}
\renewcommand{\thesection}{5.\arabic{section}}
\setcounter{theorem}{0}
\setcounter{section}{0}
 
In this chapter we establish Poincar\'e duality at the level of chainlets and cochains.   We begin by finding a chainlet representation $Ch(\o)$ of a differential form $\o$ which converts a contravariant form into a covariant chainlet.    This means that we will be able to do such things as pushforward differential forms under smooth mapping and define their boundaries.  We call these representatives of forms {\itshape \bfseries exterior chainlets}.  We have to exercise caution here.  Operators closed in the chainlet spaces may not preserve subspaces of exterior chainlets.  For example, the boundary of an exterior chainlet is not generally an exterior chainlet itself but is a chainlet of lower dimension.  The pushforward of an exterior chainlet is not generally an exterior chainlet unless the pushforward mapping is a diffeomorphism. (See Theorem \ref{change2}.)   

This leads us to a definition of cap product and a formulation of Poincar\'e duality at the level of chainlets and cochainlets  which passes to the classical duality for homology and cohomology classes.  
 
  \section*{Exterior chainlets}  
\subsection*{Inner products of $k$-elements}   Suppose $\a$ is a $k$-element   supported in a point $p.$    Since $\a$ is a  chainlet, the star operator applies.  Moreover, $\star \a$ is also a discrete  $(n-k)$-cell determined by the mass of $\a$ and star of the $k$-direction of $\a$. 
This leads to a natural definition of inner product of $k$-elements $\a$ and $\b$, supported in the same point.
$$<  \a,\b>vol:=   \a \wedge  \star \b.$$  

\begin{lemma}
$<\a,\b>$ is an inner product.
\end{lemma}

\begin{proof}  Bilinearity follows from linearity of star and bilinearity of wedge product.   Symmetry follows by symmetry of starred wedge product $\a \wedge \star \b = \b \wedge \star \a.$ By definition of star we know
$\a^*\a = vol.$ Therefore for a unit $k$-vector $\a$, $<\a,\a>vol = vol$ implies $<\a,\a> = 1.$  If $\a$ = 0 then $<\a,\a>vol = 0$ implies $<\a.\a> = 0.$  $<\a,\b>vol = \a \wedge \star \b = \b \wedge \star \a = <\b,\a>vol.$   
  \end{proof}
 
A differential  $k$-form $\o$ of class $B^r$    determines a unique  $k$-chainlet  $Ch(\o)$ of class $N^r$, called an {\itshape \bfseries exterior chainlet}.  We show below that   $$\int_M <\eta,  \o> vol =  \int_{Ch(\o)} \eta$$ 
 for all $k$-forms $\eta$ of class $B^r$.   We construct  $Ch(\o)$ as a limit of polyhedral chains.   
  
 \subsection*{Chainlet representations of differential forms}   Let $\o$ be a differential $k$-form of class $B^1$ defined in an open set $U \subset \R^n$.   Assume $U$ is bounded.  At each $p$ the form $\o$ determines a unique $k$-element $Vec(\o,p)$ via the inner product on $k$-element chains at a point.   That is, by the Riesz representation theorem there exists a unique $Vec(\o,p)$ such that
 $$\o(p,\a) = <Vec(\o(p)),\a>$$ for all $k$-vectors $\a.$      Take a cube $Q \subset U$ from the binary lattice and subdivide it into smaller binary cubes $Q_{k,i}$ with midpoint $p_{k,i}$  and edge $2^{-k}$.  Define $$\dot{P}_k = \sum_i M(Q_{k,i}) Vec( \o( p_{k,i})).$$  It is a straightforward exercise to show that  $\dot{P}_k$ forms a Cauchy sequence in the $1$-natural norm.   Denote the chainlet limit by $Ch(\o,Q)$.   Now use a Whitney decomposition to subdivide $U$ into cubes $U = \cup_{j=1}^{\i} Q^j$. Define $$Ch(\o) := \sum_{j=1}^{\i} Ch(\o, Q^j).$$  This converges in the $1$-natural norm since the total mass of the nonoverlapping binary cubes is bounded. 
 
\subsection*{Remark}  It is worth noting that a parallel attempt to show the discrete form $\dot{\o}_k = \sum_i M(Q_{k,i})  \o( p_{k,i})$ converges in the $B^r$ norm will fail.   The sequence is not Cauchy as long as the supports are finite.  However, these important and useful forms arise naturally in chainlet geometry as the dual spaces to subspaces of element chains.   Here is a place where chains and cochains are quite different.

\begin{theorem} \label{Chtheorem} Suppose $\eta, \o$ are $k$-forms.  Then
$$\int_M <\eta, \o> dV = \int_{Ch(\o)} \eta.$$
\end{theorem}

\begin{proof}
It suffices to work with the discrete approximators to $Ch(\o)$ in a cube $Q$ for both integrals.
$$
\begin{aligned} \int_{M(Q^j_{k,i}) Vec(\o(p_{k,i}))} \eta &= \sum_i M(Q^j_{k,i}) \eta(p_{k,i}) \cdot Vec(\o(p_{k,i})) \\& =  \sum_i M(Q^j_{k,i})< Vec(\eta(p_{k,i})), Vect(\o(p_{k,i}))> 
\\&= \sum_i\int_{Vec(Q^j_{k,i})} <\eta, \o> dV. 
\end{aligned}
$$

Hence
$$
\begin{aligned}
\int_{Q^j} <\eta, \o> dV &= \lim_{k \to \i} \int_{\sum_i Vec(Q^j_{k,i})} <\eta, \o> dV \\&=   \int_{Ch(\o, Q^j)} \eta.
\end{aligned}
$$
Now take write $U$ as a sum of cubes in the Whitney decomposition to obtain $$\int_U <\eta, \o> dV = \int_{Ch(\o)} \eta.$$  Since the result holds in each coordinate chart, it is valid over a manifold $M$.

 \end{proof}

\begin{corollary} Suppose $M$ is compact and $r \ge 1$.  Then there exists a constant $C > 0$ such that
$$|Ch(\o)|^{\natural_{r}} \le C|\o|_r.$$ for all $k$-forms $\o$ of class $B^r.$  
\end{corollary}

\begin{proof}
This reduces to the following estimate:
$$\left| \int_M <\eta, \o> dV \right| \le |M|^{\natural_{r}} |<\eta, \o> dV |_r \le C |\eta|_r|\o|_r$$ where $C = vol(M).$
\end{proof}

\begin{theorem}\label{dense}
$Ch: \cal{B}_k^r(M) \to \cal{N}_k^r(M)$ is a one-one linear mapping of $k$-forms of class $B^r$ into $r$-natural $k$-chains with dense image.
\end{theorem}

\begin{proof}
The mapping $\o \mapsto Ch(\o)$ is clearly  one-one and linear.  To show the image is dense, it suffices to show that for any cell $\s$ and $\e > 0$ there is an $\o$ such that
$$\left|\int \phi(X) \cdot \o - X \cdot \s\right| \le |X|^{\natural_{r}} \e$$ for any $X$ of class $N^r$. This yields $$|Ch(\o) - \s|^{\natural_{r}} < \e.$$  
Choose an $(n-k)$-cell $Q'$ through $q_0 \in \s$ orthogonal to $\s$.  The simplexes $\s(q) = T_{q-q_0} \s$ with $q \in Q'$ form a cell $Q$.  Choose $Q'$ so small that $|\s(q) - \s|^{\natural_{r}} < \e/2, q \in Q'.$  Define $\b = Vec(\s)/M(Q)$in $Q$ and $\b = 0$ in $\R^n -Q$ it follows that
$$\left|\int \phi(X) \cdot \b - X \cdot \s\right| = \left|\int_{Q'} X \cdot (\s(q) - \s)\right| \le |X|^{\natural_{r}} \e/2.$$ Choose a smooth form $\a$ in $\R^n$ so that $\int \b -\a< \e/2.$

\end{proof}

\begin{proposition}\label{eta} If $\o$ and $\eta$ are $k$-forms of class $B^r$ then
$$\int_{Ch(\o)} \eta = (-1)^k\int_{Ch(\eta)} \o.$$
\end{proposition}

\begin{proof}
$$\int_{Ch(\o)} \eta = \int_M <\eta, \o> dV = (-1)^k \int_M<\o,\eta>dV = (-1)^k \int_{Ch(\eta)} \o.$$
\end{proof}

The pushforward of an exterior chainlet is not generally an exterior chainlet, unless $f$ is a diffeomorphism.  
 
\begin{theorem}  Suppose $M$ is compact.  $Ch: \cal{B}_k^r(M) \to \cal{N}_k^r(M)$  satisfies
\begin{enumerate}
\item[(i)] $\star Ch(\o) = Ch(\star \o)$;
\item[(ii)] $f_* Ch(\o) = Ch(f^{-1*} \o)$ if $f:M\to N$ is a diffeomorphism;   
\item[(iii)] $\p Ch(\o) = Ch(\diamondsuit(\o)$;
\item[(iv)] $\d Ch(\o) = Ch(\p \o)$;
\item[(v)] $\Delta Ch(o) = Ch(\Box \o).$
\end{enumerate}

\end{theorem}

\begin{proof} (i) follows directly from the definitions of  $\star$ and $Ch$.  \\(ii)   Since $M = f^{-1}_*N$ we have
$$\begin{aligned} \int_{Ch(f^{-1*} \o)} \eta = \int_{f^{-1}_*N} <\eta,f^{-1*} \o> dV &= \int_M <f^* \eta, \o> dV \\&= \int_{Ch(\o)} f^* \eta \\&= \int_{f_* Ch(\o)} \eta.
\end{aligned}$$

 We give a proof to (v).  The remaining (i) and (iii) are similar.  
$$\begin{aligned} \int_{Ch(\Box \o)} \eta = \int <\eta, \Box \o> dV &= \int <\Box \eta, \o> dV \\&= \int_{Ch(\o)} \Box \eta \\&= \int_{\Delta Ch(\o)} \eta.
\end{aligned}
$$ The result follows.  
\end{proof}
\begin{conjecture} A chainlet $A$ of class $N^r$ is harmonic iff $A = Ch(\o)$ for some  harmonic $\o$ of class $B^r$.
\end{conjecture}

\section*{Cap product}
 
 We have seen that $k$-vectors are in one-one correspondence with $k$-elements.  A $k$-vector $\a$ is also a  $k$-covector via the inner product.  If we wish to consider $\a$ as a covector, we will denote it by $\widetilde{\a}.$
 If $\a$ is a $k$-element and $\b$ is a $j$-element, we can define several products.  If $j = k$ we have the {\itshape \bfseries scalar product}
   $$\widetilde{\a} \cdot \b := <\a,\b>.$$  
   
\begin{lemma}
$\star \widetilde{ \a}   \cdot \b = \widetilde{\a} \cdot \star \b.$
\end{lemma}

    If  $ j < k$ then $\widetilde{\a} \lfloor \b$ denotes the $(k-j)$-covector
  $$(\widetilde{\a} \lfloor \b) \cdot \g :=  <\a ,  \b \wedge \g>,$$
   for all  $(k-j)$-elements  $\g$, and is called the {\itshape \bfseries interior product} of $\widetilde{\a}$ with $\b$.  This can be seen geometrically as the orthogonal complement of $\b$ in $\a$, i.e., $\b \wedge \nu = \a$ and $\widetilde{\nu} = \widetilde{\a} \lfloor \b.$
   
     If $j < k$ then $\widetilde{\a} \cap \b$ is the unique $(k-j)$-element given by the Riesz representation theorem
$$\eta \cdot  (\widetilde{\a} \cap \b) := < \eta \wedge \a,  \b>.$$  This is called the {\itshape \bfseries cap product} of  $\widetilde{\a}$ with $\b$.  A geometrical definition can be given as the $(k-j)$-element   $\nu$ orthogonal to $\a$ so that $\nu \wedge \a = \b.$

 A differential  $(j+s)$-form $\o$ of class $B^r$ and a $j$-cochain $X$ of class $N^r$ determine a product, also called {\itshape \bfseries cap product},
 $$X \cap Ch(\o): = Ch(\eta)$$ where $\eta(p) := \phi(X)(p) \cap \o(p).$

\begin{lemma}\label{XCH}
$$ \int_{X \cap Ch(\o)} \phi(Y) =  \int_M <\phi (Y) \wedge \phi (X), \o> dV.$$
\end{lemma}

\begin{proof}
Defining $\eta$ as above it follows that  
$$
\begin{aligned} \int_{X \cap Ch(\o)} \phi(Y) &= \int_{Ch(\eta)} \phi(Y) \\&= \int_M <\phi(Y), \eta> dV \\&= \int_M <\phi (Y) \wedge \phi (X), \o> dV.
\end{aligned}
$$

\end{proof}

\begin{lemma}\label{lemdot}  Suppose  $X$ is a $j$-cochain and $Y$ is a $k$-cochain of class $N^r$ with $j < k$.  Then 
$$ (Y \cup X) \cdot Ch(\o)=  Y \cdot (X \cap Ch(\o)).$$
\end{lemma}   

\begin{proof}  It follows from the definition of cap product that 
$$<\phi(Y)(p) , \phi(X)(p) \cap \o(p)> = <\phi (Y)(p) \wedge \phi( X)(p), \o(p)>.$$ Taking integrals we have
$$\int_M <\phi(Y)(p) , \phi(X))(p) \cap \o(p)> dV = \int_M <\phi Y(p) \wedge \phi X(p), \o(p)>dV.$$

  By  Theorem \ref{Chtheorem} and Lemma \ref{XCH} it follows that
$$\begin{aligned} (Y \cup X) \cdot Ch(\o)  &= \int_{Ch(\o)} \phi(Y) \wedge \phi(X) \\&=  \int_M <\phi (Y) \wedge \phi (X), \o> dV \\&=  \int_{X \cap Ch(\o)} \phi(Y) \\& = Y \cdot (X \cap Ch(\o)).\end{aligned}$$

\end{proof}

\begin{lemma}\label{Choineq}  Suppose $X$ is a $j$-cochain and $\o$ is a $k$-form, $j < k$.  Then 
$$|X \cap Ch(\o)|^{\natural_{r}} \le C|X|^{\natural_{r}}|Ch(\o)|^{\natural_{r}}.$$
\end{lemma}

\begin{proof}  Suppose $Y$ is a $(k-j)$-cochain.  Then by the Fundamental Integral Inequality of chainlet geometry 
 $$
 \begin{aligned}|Y \cdot (X \cap Ch(\o))| &= |(Y \cup X) \cdot Ch(\o)| \\&\le |Y \cup X|^{\natural_{r}}|Ch(\o)|^{\natural_{r}} \\&\le C|Y|^{\natural_{r}}|X|^{\natural_{r}}|Ch(\o)|^{\natural_{r}}. \end{aligned}$$    It follows that $$|X\cap Ch(\o)|^{\natural_{r}} \le C|X|^{\natural_{r}}|Ch(\o)|^{\natural_{r}}.$$

\end{proof}

Let $A$ be a chainlet of class $N^r$.  By \ref{dense} $A = \lim Ch(\o_i).$   Define 
$$X \cap A := \lim X \cap Ch(\o_i).$$ This limit exists as a consequence of Lemma \ref{Choineq} and yields
 
\begin{theorem}\label{capcont} For each $r \ge 1$ there exists a constant $C>0$ such that
 if $X$ is a $p$-cochain and $A$ is a  $(p+q)$-chainlet of class $N^r$  then $$|X \cap A|^{\natural_{r}} \le C|X|^{\natural_{r}}|A|^{\natural_{r}}.$$
\end{theorem}

\begin{theorem} If $X$ is a $p$-cochain $Y$ is a $q$ cochain and $A$ is a $(p+q)$-chainlet of class $N^r$ then \quad \\ \quad
\vspace{-.2in}
\begin{enumerate}
\item[(i)] $(X \cup Y) \cdot A = X \cdot (Y \cap A).$
\item[(ii)] $X \cap (Y \cap A) = (X \cup Y) \cap A$;
\item[(iii)] $\p (X \cap A) = (-1)^{p+1} dX \cap A + X \cap \p A$;
\item[(iv)] $I^0 \cap A = A;  I^0 \cdot (X \cap A) = X \cdot A$ where $I^0$ denotes the unit $0$-form $I^0(p) \equiv 1$.
\item[(v)] $f_*(f^*X \cap A)= X \cap f_* A.$
 
\end{enumerate}
\end{theorem}
 
\begin{proof}

\begin{enumerate}
   \item[(i)] First approximate $A$ with exterior chainlets $A = \lim Ch(\o_i).$  Then apply    Lemma \ref{lemdot} and Theorem \ref{capcont}.
 \item[(ii)]By (i)  if $Z$ is a cochain of class $N^r$ then
 $$
 \begin{aligned} Z \cdot (X \cap (Y \cap A)) &= (Z \cup X) \cdot (Y \cap A) \\&= ((Z \cup X) \cup Y) \cdot A \\&= (Z \cup (X \cup Y)) \cdot A \\&= Z \cdot ((X \cup Y) \cap A).
 \end{aligned}
 $$  It follows that $X \cap (Y \cap A) =  (X \cup Y) \cap A.$
 \item[(iii)] By (i) and Leibnitz' rule for cochains
 $$
 \begin{aligned} Y \cdot (\p (X \cap A)) &= dY \cdot (X \cap A)
 \\&= (dY
 \cup X) \cdot A  \\&= (d(Y \cup X) + (-1)^{p+1} (Y \cup dX))
 \cdot A   \\&= (Y \cup X) \cdot \p A + (-1)^{p+1} (Y
 \cup dX) \cdot A \\&= Y\cdot (X \cap \p A + (-1)^{p+1} dX
 \cap A).
 \end{aligned}$$
 \item[(iv)] $Y \cdot (I^0 \cap A) = (Y \cup I^0) \cdot A =  Y \cdot A; \mbox{ and }   I^0\cdot (X \cap A) = (I^0 \cup X) \cdot A = X \cdot A.$
 \item[(v)] 
 $$
\begin{aligned}
Y \cdot f_*(f^* X \cap A) &= f^*Y \cdot f^*X \cap A \\&= (f^*Y \cup f^* X) \cdot A \\&= f^* (Y \cup X) \cdot A \\&= Y \cup X \cdot f_* A \\&= Y \cdot (X \cap f_* A)
\end{aligned}
$$ 
 
  \end{enumerate}

\end{proof}

 \section*{Poincar\'e duality of chains and cochains}  An
 oriented compact $n$-manifold $M$ corresponds to a unique $n$-chainlet 
 $A_M$ in the sense that integrals of forms coincide
 $\int_M
 \o = \int_{A_M}\o.$  The cap product determines a
 Poincar\'e duality homomorphism
 $$PD: ({\cal N}_p^{r})^{\prime}
 \to  {\cal N}_{n-p}^{r}$$ by $$PD(X) := X \cap
 A_M.$$  
 
\begin{theorem}  If $\o$ is a $k$-form of class $B^r$ then
$$\phi(\o) \cap M = Ch(\star \o).$$
\end{theorem}

\begin{proof} Suppose $\eta$ is a $(n-k)$-form of class $B^r$.  By \ref{XCH} $$\int_{\phi(\o) \cap M} \eta = \int_M <\eta, \o> dV = \int_M \eta \wedge \star \o = \int_{Ch(\star \o)} \eta.$$

\end{proof}
It follows from \ref{eta} that
 $$X \cdot Ch(\o) = (-1)^k\int_{X \cap M} \o.$$
 
As another consequence, we have $$PD(X) = Ch(\phi(\star X)).$$
  
 The Poincar\'e duality homomorphism is not an
 isomorphism, although its image is dense in the space of chainlets. 
 Consider a $1$-cell
 $Q$ that is a straight line segment.  From the
 isomorphism theorem we know that if $\o \in {\cal
 B}_p^{r,\a}$  then
 $$PD(\Psi(\o))(\nu) = \int_M \o \wedge\nu $$ for all $\nu \in
 {\cal B}_{n-p}^{r}$. Therefore the inverse would have
 to an $r$-smooth differential form $\o$ satisfying 
 $\int_Q \nu = \int_M \o \wedge \nu$ for all
 $r$-smooth $\nu$.   However, the only possibility
 would be a form defined only on $A$.        
 
 A cochain $X$ is {\itshape \bfseries harmonic} if
 $\Box X = 0.$   If $X$ is a cocyle then it is harmonic if $d \d X = d \star d \star X = 0.$  A chainlet $A$ is {\itshape \bfseries harmonic} if $\Delta A = 0.$     If  $A$ is a cycle it is harmonic 
 if $ \p \diamondsuit A =  \p\star \p \star  A  = 0.$
  This extends the previous definition
 of $PD$ defined on cochains.
 \begin{theorem}\label{PD} $$PD: ({\cal N}_p^{r})^{\prime}
 \to  {\cal N}_{n-p}^{r}$$ is a homomorphism satisfying
 $$PD(dX) = (-1)^{p+1}\p PD(X)$$
 $$PD(\star X) = \star PD(X)$$
 $$PD(\d X) = \pm \diamondsuit PD(X);$$
 $$PD(H) \mbox{ is harmonic if }  H \mbox{ is harmonic and $M$ is closed.}$$
 \end{theorem}

 If $X$ and $Y$ are cocycles with $X - Y = dZ$ then $PD(X)$ and $PD(Y)$ are cycles and $PD(X) - PD(Y) = PD(dZ) = \p PD( (-1)^{p+1}Z). $ It follows that $PD$ passes to singular cohomology and homology classes.   Since classic Poincar\'e duality is an isomorphism of singular cohomology and homology classes, then $PD$ is also an isomorphism at this level since the definitions coincide.

  This theorem also implies that PD preserves Hodge decompositions since coboundaries are sent to coboundaries, boundaries are sent to boundaries and harmonic chainlets are sent to harmonic chainlets.   Let $\cal{E}$ denote the dense subspace of chainlets of the form $Ch(\o).$  We know that $\cal{E}$ is closed under the operators of $\p$ and $\star.$
\begin{corollary}
 If $A \in \cal{E}$ then there exist $B, C, H  \in \cal{E}$ such that
  $$A = \p B + \diamondsuit C + H$$ where  $H$  is  a harmonic chainlet.
\end{corollary}   

 Observe that this result is not possible via either the sharp or flat topologies of Whitney where harmonic chains are not defined.   Either the star operator or boundary operator is missing.  (See Table 1 of the preface.)  
   It should be of considerable interest to 
study the spectrum of the   geometric Laplace operator $\Delta$ on chainlets.  This is also important for the development of chainlet Hodge theory. The geometric Hodge and Laplace operators were originally defined by the author for the purpose of developing a geometric Hodge theory for chainlets.   With the extension of chainlet geometry to Riemannian manifolds developed in these notes, and the isomorphism theorem of \cite{currents} between summable currents and chainlets extended to compact Riemannian manifolds $M$, a  Hodge decomposition for chainlets is immediate. \footnote{The unpublished  1998 Berkeley thesis of J. Mitchell, initially drafted under the supervision of the author and completed with Morris Hirsch, attempted to do this, but not all tools were available at the time.  In particular, chainlets were not defined on Riemannian manifolds and the isomorphism between currents and chainlets was not well understood \cite{currents}. }
\begin{theorem} If $C$ is a chainlet in a compact Riemannian manifold $M$ there exist unique chainlets $\p A, \d B$ and $H$ in $M$, where $H$ is harmonic, such that $$C = \p A + \delta B + H.$$
\end{theorem}  
\begin{proof} (sketch)
The chainlet  $C$ determines a unique summable current $c$ via $c(\o) = \int_C \o.$  The current has a unique Hodge decomposition 
$c = \p a + \d b + h$ where $a, b, h$ are currents of the appropriate dimension.    Each is associated to a unique chainlet $A, B, H$. The operators are respected under the operations of $\star$ and $d$.  The result follows. 
\end{proof}

 This is all well and good, but we would like to see a Hodge decomposition of chainlets for all Riemannian manifolds and constructed  geometrically.  We would also like to see   methods for calculations and specific examples. There is much remaining to do.  Some of this will be treated in more detail in the second part of these lecture notes.     
 
 \section*{Inner products}

Since every Hilbert space is reflexive, we know that chainlet spaces are not  Hilbert spaces.  However, we can identify a dense subspace  of chainlets for which there is an inner product defined.

\subsection*{Effective Hilbert spaces} 
The author calls a Banach space $X$ with norm $|\quad|$  an {\itshape \bfseries effective Hilbert space} if there exists a dense vector subspace $Y \subset X$ and an operator $<\quad, \quad>: X \times Y \to \R$ such that 
\begin{enumerate}
\item[(i)] $<\quad, \quad>$ is an inner product on $Y \times Y$;
\item[(ii)] $\sqrt{<B, B>} = |B|$ for all $B \in Y$;
\item[(iii)] $<u +v,w> = <u,w> + <v,w>$ for all $u,v \in X, w \in Y$, 
$<w, u+v> = <w,u> + <w,v>$ for all $w \in X, u,v \in Y$; 
\item[(iv)] $<au,v> = a<u,v> = <u, av>$ for all $u \in X, v \in Y, a \in \R.$ 
\end{enumerate}

We show that the Banach space $\cal{N}^{\i}$ of chainlets is an effective Hilbert space and find subspaces $Y$ and  inner products that satisfy the conditions of the definition.  
First set  $Y = \cal{E}$,  the subspace of exterior chailnets of $\cal{N}^{\i}$ consisting of chainlets of the form $X \cap M$ where $X \in \cal{N}^{\i \prime}$ is a cochain.  

\begin{lemma} If $M$ is compact, $$|X \cap M|^{\natural_{r}} \le vol(M) |X|^{\natural_{r}}.$$
\end{lemma}

\begin{proof}
$$|X \cap M|^{\natural_{r}} = |Ch(\phi(X)|^{\natural_{r}} \le vol(M) |X|^{\natural_{r}}.$$

\end{proof}

\setcounter{definition}{0}
\begin{definition}{Inner product on $\cal{N} \times \cal{E}$}
$$<A, X \cap M>:= |\star X \cdot    A|.$$
\end{definition} 

\begin{proposition}
$<\p A,   Ch(\o)> = (-1)^k<A , \diamondsuit Ch(\o)>.$
\end{proposition}

Define $\|E\| = \sqrt{<E,E>}$
\begin{theorem}
$\|E\| \sim |E|^{\natural_{}}.$
\end{theorem}

Remark:
$$<Ch(\a), Ch(\b)> = <\a,\b>.$$

 \chapter{Locally compact abelian groups} \begin{center} \em{\large \color  [gray]{0.5} DRAFT}
 \end{center} 
 \renewcommand{\thetheorem}{6.\arabic{theorem}}
\renewcommand{\thesection}{6.\arabic{section}}
\setcounter{theorem}{0}
\setcounter{section}{0}
 We recall that a topological group is {\itshape \bfseries locally compact} if and only if the identity $e$ of the group has a compact neighborhood.   Let $G$ by a locally compact abelian group.   Translation $T_v$ through a vector $v$ is replaced by translation through a group element $g$. That is, $T_g(U) = U + g$ where $U \subset G.$   (We write our group action as addition rather than the more standard multiplication simply to fit with our previous exposition.  This is strictly notational.)
 For our norms to carry through, we assume $k=0$ so that cells are all $0$-dimensional.  
 
 The cells $\g$ of $G$ will be sufficiently regular elements of the $\s$-algebra ${\cal A}(G)$ generated by the compact subsets.  It is remarkable that there exists an essentially unique natural measure, the {\itshape \bfseries Haar measure}, that allows us to measure the elements of ${\cal A}(G).$  Haar measures are unique up to positive scale factors.  
 More precisely, a right Haar measure $\mu$ on a locally compact group $G$ is a countably additive measure   defined on the Borel sets of $G$ which is right invariant in the sense that $\mu(A + x) = \mu(A)$ for $x$ an element of $G$ and $A$ a Borel subset of $G$ and also satisfies some regularity conditions.   For $x \in G$, denote $|x| = \mu(x).$  (We distinguish single elements $x \in G$ and elements of ${\cal A}(G)$ to fit with our previous exposition. Certainly $x \in {\cal A}(G)$.)  
   
\subsubsection*{Examples of locally compact abelian groups}  

\begin{itemize}
\item $\R^n$ with vector addition as group operation.

\item The circle group $T$. This is the group of complex complex numbers of modulus 1. $T$ is isomorphic as a topological group to the quotient group $\R/\Z.$ 
\end{itemize}

Haar measure allows to define the notion of integral for (complex-valued) Borel functions defined on the group. In particular, one may consider various $L^p$ spaces associated to Haar measure. Specifically, $$L_{\mu}^p(G) =\left\{f:G \to \C: \int_{G}|f(x)|^p d\mu(x)\right\}.$$

The Banach space $L^1(G)$ of all $\mu$-integrable functions on $G$ becomes a Banach algebra under the convolution $xy(g) = \int x(h) y(h^{-1}g) d\mu(h)$ for $x, y \in L^1(G)$.

 If $G$ is a locally compact abelian group, a {\itshape \bfseries character} of $G$ is a continuous group homomorphism from $G$ with values in the circle group $T$. It is well known that the set of all characters on $G$ is itself a locally compact abelian group, called the {\itshape \bfseries dual group} of $G$, denoted $G^{\wedge}$. The group operation on $G^{\wedge}$  is given by pointwise multiplication of characters, the inverse of a character is its complex conjugate and the topology on the space of characters is that of uniform convergence on compact sets. This topology is not necessarily metrizable. However, if the group $G$ is a 
  locally compact abelian group, then the dual group is metrizable.

\begin{theorem}
$(G^{\wedge})^{\wedge}$ is canonically isomorphic to $G$.
\end{theorem} 

The isomorphism is given by $x \mapsto \{\chi \mapsto \chi(x)\}.$
We define a {\itshape \bfseries distribution} to be an element of the dual space of $G.$  

Our next goal is to extend the definitions of the natural norm to  $G$ and obtain a theory of distributions on $G$ that   does not require the notion of differentiation.  (See \cite{feich} for a related result.)

For $g \in G$ let  $T_g$ denote translation through $g$
 $$T_g(A) = A + g.$$  Let   $\g^0 \in \cal{A}(G)$    and $g_1, \cdots, g_r \in G$.    Define  $$\g^1 = \g^0 - T_{g_1}\g^0$$  and   $$\g^{j+1} = \g^j  - T_{g_{j+1}} \g^j.$$
 
  Let
$$D^j =
\sum_{i=1}^m a_i \g_i^j$$
with coefficients $a_i \in \Z.$ This is called a $G$-chain of order $j$. The vector space of all such $G$-chains $D^j$ is
denoted  $\cal{D}^j.$      A $G$-chain $D^0$  of order 0 is sometimes simply called a $G$-chain.

\subsubsection*{$G$-chain mass} Given  $\g^j$   generated by $\g^0 \in G$ and $g_1,  \cdots , g_j \in G$, define $M(\g^0) = \|\g^0\|_0 = \mu(\g^0)$ and for $j \ge 1$, 
$$\|\g^j\|_j = \mu(\g^0)|g_1| |g_2| \cdots |g_j |.$$
 For $D^j = \sum_{i=1}^m a_i \g_i^j$, possibly overlapping, define $$\|D^j\|_j = \sum_{i=1}^m |a_i|\| \g_i^j\|_j.$$

\subsubsection*{Natural $G$-norms} 
 
For $r \ge 0$ define the {\itshape \bfseries r-natural} $G$-norm
$$|P|^{\natural_r}  = \inf\left\{\sum_{j=0}^r\|D^j\|_j  \right\}$$  
where the infimum is taken over all decompositions
$$P = \sum_{j=0}^r D^j  $$
where $D^j \in \cal{D}^j.$     It is clear $|\quad|^{\natural_r}$ is a semi-norm. 
 We shortly prove it is a norm.
 
  Define $$\int_{\g}f := \int_G \chi_{\g} f d\mu.$$

  Suppose $f:G \to T.$  Define $$\|f\|_0 := \sup \left\{\frac{|\int_{\g} f |}{\mu(\g)}: \g \in \cal{A}(G) \right\}.$$ 
  Inductively define $$\|f\|_r := \sup\left\{\frac{\|f -T_vf\|_{r-1}}{|v|}\right\}.$$  
 
Define $$|f|_0 := \|f\|_0$$ and  for $r \ge 1$,
  $$|f|_r := \max\{\|f\|_o, \cdots, \|f\|_r \}.$$
   We say that $f$ is of class $B^r$ if $|f|_r < \i.$  Let $ \cal{B}_G^r$ denote the space of functions on $G$ of class $B^r.$

Given a compact subset $K \subset M$, there exists a smooth  function $f_K$ which vanishes outside an $\e$-neighborhood of $K$ and which is nonzero on $K$.    If $M$ is infinitely smooth, we can also find a function $\phi_K:M \to \R$ which is nonzero on $K$  and such that all the derivative of $\phi_K$ are uniformly bounded by a constant $C$.

\begin{theorem}\label{oldintegral2} Let $P \in \cal{D}^0$, $r \in \Z^+,$ and $f \in \cal{B}_G^r$ defined in a neighborhood of $spt(P).$ Then
$$\left|\int_P f \right| \le |P|^{\natural_r}|f|_r.$$
\end{theorem}
 
\begin{proof}    We first prove $ \left|\int_{\g^{j}} f \right| \le \|\g^j\|_{j}\|f\|_{j}.$
By the definition of $\|f\|_0$ we know
$$\left|\int_{\g^0} f \right| \le   \|\g^0\|_0\|f\|_0. $$    
 
 Apply induction to deduce
$$\begin{array}{rll} \left|\int_{\g^{j}} f \right| =  \left|\int_{\g^{j-1} - T_{v_j}\g^{j-1}} f \right| 
&=  \left|\int_{\g^{j-1}} f - T_{v_j}^* f \right|  \\&\le  \|\g^{j-1}\|_{j-1}\|f - T_{v_j}^*f\|_{j-1} \\&\le  \|\g^{j-1}\|_{j-1}\|f\|_{j}|v_{j}| \\&= \|\g^j\|_{j}\|f\|_{j}\end{array}$$
 
  By linearity $$\left|\int_{D^j} f\right| \le \|D^j\|_j \|f\|_j$$
for all  $D^j \in \cal{P}_G^j$. 
 
  We again use induction to prove $\left|\int_P f\right| \le  |P|^{\natural_r}|f|_r.$ 
As before,   $\left|\int_P f\right| \le |P|^{\natural_0}|f|_0.$  
Assume the estimate holds for $r-1.$ 

  Let $\e > 0$. There exists $P = \sum_{j=0}^r  D^j   $ such that $|P|^{r} >
\sum_{j=0}^r \|D^j\|_j   - \e$. By induction
 $$\begin{array}{rll}  \left|\int_P f\right| &\le \sum_{j=0}^r \left|\int_{D^j} f \right|   \\& \le \sum_{j=0}^r \|D^j\|_j\|f\|_j  \\& \le (\sum_{j=0}^r
\|D^j\|_j  ) |f|_r\\&\le  (|P|^{\natural_r} +
\e) |f|_r.

\end{array} $$
Since the inequality holds for all $\e > 0$ the result follows.
\end{proof} 
\begin{corollary} $ |P|^{\natural_r}$ is a norm on the space of $G$-chains $\cal{P}_k$. \end{corollary}
\begin{proof} Suppose $P \ne 0$ is a $G$-chain. There exists a function $f$   such that $\int_P  f \ne 0$ and $|f|_r < \i.$  Then
$0 <  \left|\int_P f\right| \le |P|^{\natural_r}|f|_r$ implies $|P|^{\natural_r} > 0.$ \end{proof}
The Banach space of     $G$-chains $\cal{P}_k$ completed with the
norm $|\quad |^{\natural_r}$ is denoted $\cal{G}^r $. The elements of $\cal{G}^r $ are called
{\itshape \bfseries $G$-chainlets of class} $N^r$.
 We let $ \cal{G}^{\i}$ be the direct limit of the $\cal{G}^r.$  Define $|A|^{\natural_{}} = \lim_{r \to \i}|A|^{\natural_{r}}.$  This is a norm since $\left|\int_A f\right| \le |A|^{\natural_{}} |f|_{\i}$ and there exists $f$ such that $\int_A f > 0$ and $|f|_{\i} < \i.$  The characterization of the Banach space is essentially the same as for chainlets, except there is no boundary operator. 
 
\begin{theorem}
The Banach space $\cal{G}^{\i}$  is the smallest Banach space containing $X^0$ and which has  Lipschitz bounded translation operators.

\end{theorem}

Since the space is reflexive, these $G$-chainlets correspond to distributions over $G$.  We obtain a theory of distributions on $G$ that  does not require the notion of differentiation.

\chapter{Discrete calculus}\label{discrete} 
\addtocounter{chapter}{1}
\renewcommand{\thetheorem}{7.\arabic{theorem}}
\renewcommand{\thesection}{7.\arabic{section}}
\setcounter{theorem}{0}
\setcounter{section}{0}

 	\begin{center} \em{\large \color  [gray]{0.5} DRAFT}
 \end{center}
 
In the Ravello lectures the author presented two discrete theories with different flavors and applications.  The first is drafted below in this first part of the lecture notes.  The second discrete theory will appear in the second part of the lecture notes, under preparation.   In both, we relax the use of  triangulations, tesselations, cubical subdivisions, or the like, and no longer require that $k$-elements be connected to one another.    Domains are supported in countably many points.   We replace connectivity with multiplicity. 
   
   The first  theory uses discrete chains and assumes that we have at our disposal classically defined smooth differential forms.      The second approach is a full discrete theory where both chains and cochains are supported in the same countable set of points and will be fully presented in the second part of these lecture notes.  
 In the full discrete theory, cochains are ``measuring sticks'' that match chains perfectly.  We characterize the cochains by what the author calls ``discrete forms''.    Indeed, we propose replacing the venerable ``Whitney forms'' with discrete forms because our class of discrete forms is closed under the operations of $\star$ and $d$.  Furthermore, cochains are associative and graded commutative.  The operators $\star, d$ and $\wedge$ work seamlessly with each other at the discrete level and converge to the smooth continuum.        
\subsection*{Differential $k$-elements}
  
  We saw in  Chapter \ref{fractals} that $k$-elements embed in the chainlet space $\cal{N}_k^1$.   The subspace of $k$-element chains is denoted $V_k^0.$   Those supported in a set $J$ are denoted $V_k^0(J).$  We shall see that there is a vast array of chainlets supported in a single point and much of calculus can be found locally by understanding these and their operators and products.

A $k$-element is a mass normalized version of a cell, shrunk to a point.   Elements are reminiscent of   Dirac monopoles, but they have direction, dimension, multiplicity and orientation.   We next extend the notion of Dirac dipoles, working towards a  Grassman algebra of differential elements.  

Let $\dot{\s}(p)$ be a $k$-element   with $k$-direction $\a$ and supported at $p$.   We define its {\itshape \bfseries geometric directional derivative}.  Choose a tangent vector $v$ based at $p$.  Consider the sequence of {\em difference elements}   
$$\dot{Q}_i = 2^i(\dot{\s}(p) - T_{v/2^i}\dot{\s}(p)), i \ge 0.$$  This forms a Cauchy sequence in the $2$-natural norm since
$$|\dot{Q}_i - \dot{Q}_{i+j}|^{\natural_{2}}  \le \|\dot{Q}_i - \dot{Q}_{i+j}\|_2 \le 2^{-i} M(\dot{\s}(p)).$$    (See Corollary \ref{natnorms}.) 

Remark. In order to make the notion of mass of a $k$-chainlet precise,  we need to establish lower semi-continuity of mass in the natural norms.  This will be addressed in the second part of these lecture notes.   

The author calls the limit of the $\dot{Q}_i$ a {\itshape \bfseries differential $k$-element of order $1$} and denotes it by $\nabla_v \dot{\s}(p).$  It follows immediately from the definition of the operator $\nabla_v$ that $$\int_{\nabla_v \dot{\s}(p)}  \o = \int_{\dot{\s}(p)} \nabla_v \o.$$
The definition of directional derivative extends linearly to  $k$-element chains $\dot{P} = \sum a_i \dot{\s_i}$, yielding $\nabla_v \dot{P} = \sum a_i \nabla_v \dot{\s}_i.$   
  Hence
\begin{theorem}  If $\o$ is a differential form of class $B^2$ and $\dot{P}$ is a $k$-element chain then
$$\int_{\nabla_v \dot{P}}  \o = \int_{\dot{P}} \nabla_v \o$$  
 and
$$|\nabla_v \dot{P}|^{\natural_{2}} \le |\dot{P}|^{\natural_{1}}.$$

\end{theorem}

\begin{proof}
$$|\nabla_v \dot{P}|^{\natural_{2}} = \sup \frac{|\int_{\nabla_v \dot{P}}  \o|}{|\o|_2}  \le \sup \frac{|\int_{\dot{P}} \nabla_v \o|}{|\nabla_v \o|_1} = |\dot{P}|^{\natural_{1}}.$$
\end{proof}

We may therefore define the directional derivatives of chainlets by taking limits of element chains $\dot{P}$ in the $1$-natural  norm:  $$\nabla_v: \cal{N}_k^1 \to \cal{N}_k^2$$ and deduce 

\begin{theorem}  If $\o$ is a differential form of class $B^2$  and $A$ is a chainlet of class $N^1$ then
$$\int_A \nabla_v \o = \int_{\nabla_v A}\o.$$
\end{theorem}

\section*{Lie derivatives of element chains}
If $X$ is a vector field and $ \a$ is a field of $k$-vectors    we obtain a Lie derivative of $\a$ in the direction $X$.  Let $f_t$ be the flow of $X$.  Fix $x_0$ and define $$\cal{L}_X (\a)_{x_0} := lim_{t \to 0} \frac{f_{-t*} \a_{x_t} - \a_{x_0}}{t}.$$     If $X$ is smooth then the limit exists.   

We similarly define $\cal{L}_X$ on fields of differential $k$-elements or order $s$.   The limit exists in the $s$-natural norm as a field of differential $k$-elements or order $s$.    

\begin{lemma}
$$\cal{L}_X \p  = \p \cal{L}_X.$$
\end{lemma}
\begin{proof}
$$\cal{L}_X \p \a = \lim \frac{f_{-t} \p \a_{x_t} - \p \a_{x_0}}{t} = \p \lim\frac{f_{-t}  \a_{x_t} -  \a_{x_0}}{t} = \p \cal{L}_X \a.$$
\end{proof}

Define $$\cal{L}_X \o(p;\a) := \o(p; \cal{L}_X \a).$$  

\begin{proposition}
The above definition coincides with the standard definition of Lie derivative of a differential form.  Furthermore, $$d \cal{L}_X = \cal{L}_X d.$$
\end{proposition}

\begin{proof}
$$d \cal{L}_X \o(p; \a) = \cal{L}_X \o(p; \p \a) = \o(p; \cal{L}_X \p \a) = \o(p; \p \cal{L}_X \a) = \cal{L}_X d\o(p; \a).$$ 
\end{proof}

 \section*{An equivalent discrete norm}  The notion of a difference $k$-cell of order $i$ naturally extends to all $k$-chainlets.  Let $\dot{\s}_i$ denote a difference $k$-element of order $i$.  That is, there exists a $k$-element  $\dot{\s_0}$ and vectors $v_0, \cdots, v_i$ such that $\dot{\s}_i = (Id -T_{v_i}) \circ \cdots \circ (Id - T_{v_0}) \dot{\s}_0.$  Let $\dot{D}^i = \sum a_j \dot{\s}^i_j$ denote a difference $k$-element chain of order $i$ and $\|\dot{D}^i\|_i = \sum_j |a_j| \|\dot{\s}^i_j\|_i.$  Denote the vector space of all such difference $k$-element chains of order $i$ by $\dot{\cal{D}}_k^i$.

\begin{theorem}   Let $\dot{P} \in V_k^0.$  Then 
$$|\dot{P}|^{\natural_{r}} = \inf\left\{\sum_{i =1}^r \|\dot{D}^i\|_i: \dot{P} = \sum_{i = 1}^r \dot{D}^i ,  \dot{D}^i \in \dot{\cal{D}}_k^i\right\}.$$ 
\end{theorem}

\begin{proof} To establish $\le$ this reduces to showing that  
 $\|\dot{D}^i\|^{\natural_{r}} \le \|\dot{D}^i\|_i$. This follows from Corollary \ref{natnorms}.  For the other direction, we again use Corollary \ref{natnorms}.   
 $$|\dot{P}|^{\natural_{r}} = \inf\left\{\sum\|D^i\|_i + |C|^{\natural_{r}}: \dot{P} = \sum D^i + \p C\right\}$$ where $D^i $  is a difference $k$-chainlet chain of order $i$ and $C$ is a $(k+1)$-chainlet of class $N^{r-1}.$   Thus for $\e > 0$, there exists $\dot{P} = \sum D^i + \p C$ such that 
 $$|\dot{P}|^{\natural_{r}} >  \sum\|D^i\|_i + |C|^{\natural_{r-1}} - \e/2.$$  We may assume that the rhs consists of difference element chains since the lhs is an element chain.    Furthermore, $C$ must be a  difference element chain, $C = \sum \dot{E}_i. $
 By induction $$|C|^{\natural_{r-1}} \ge \sum \|\dot{E^i}\|_i - \e/2.$$     (This is immediate for $r =1.$ )
   Thus 
$\dot{P} = \sum \dot{D^i}  + \sum \dot{E}^i $ and $$|\dot{P}|^{\natural_{r}} > \sum\|\dot{D^i}\|_i + \sum\|\dot{E^i}\|_i - \e.$$
\end{proof} 
 
We remark that one cannot omit the boundary term in the polyhedral definition of the natural norms and obtain an equivalent norm.   Without it, we would not be able to prove that staircases converge to the diagonal.  However, with elements, shape is no longer important.  Discrete staircases do converge to discrete diagonals without a boundary term being  used.

\subsection*{Higher order differential elements}$\quad$
Define {\itshape \bfseries   differential $k$-elements of order $s$} as   the directional derivatives of  differential $k$-elements of order $s-1.$

\subsection*{Boundaries of $k$-elements} The boundary operator $$\p : V_k^s \to V_{k-1}^{s+1}$$ sends a  differential $k$-element chain of order $s$   to a  differential $(k-1)$-element chain of order $(s+1)$.  Since $k$-element chains are dense in chainlets of class $N^r$, $r \ge 1$, linear functionals on them correspond to differential forms of class $B^r$, as was the case for polyhedral chains.  One may therefore build a discrete theory using $k$-element chains as approximators to domains.  However,   it turns out to be more efficient and concise to place   differential $k$-element cells of order $s$ on an equal, but graded, footing with $k$-element cells and not be limited to finding them by taking limits of chains of the latter.   Let the vector space of $k$-element chains of order $s$ be denoted by $V_k^s$ and define $$V_k := V_k^0 \oplus V_k^1 \oplus \cdots.$$   
\subsection*{Pushforward of $k$-elements}
The pushforward operator $$f_*: V_k \to V_k$$  sends $k$-cells of order $s$ to $k$-cells of order $s$ via the total derivative at the supporting point.  It allows dynamical systems to be modeled as we may replace $f$ by the time-t map of a flow $f_t$.   Each diffeomorphism $f_t$ becomes linear at a point and takes the form $f_{t*}(p, \a) = (f(p), f_*(\a))$

\begin{theorem}
$$f_* \p = \p f_*.$$
\end{theorem}

\subsection*{Example}(Dedicated to my Ravello friends)  A vortex, such as a hurricane, is approximated by time parametrized $k$-element chains $$\sum f_{t*}(p_i, \a_{p_i})$$ where each $f_{t*}$ is a linear rotation.  This is related to the vortex models of Chorin, but the full calculus is established for our models, including the boundary operator, star operator and the divergence theorem.

Similar examples exist for arbitrary dynamical systems, including saddles, sinks, sources, chaotic horseshoes, and the like.
\subsection*{Operators on differential forms} At this point we summarize new definitions of operators on differential forms.  The goal is to define as much as possible at a geometrical level and obtain operators on forms through duality.

\begin{itemize}
\item $f^* \o(p; \a) =\o(f(p); D^kf_p(\a))$
\item $\star \o(p; \a) := \o(p; \star \a)$
\item $d \o(p; \a) := \o(p; \p \a)$
\item $\nabla_v \o(p; \a) := \o(p; \nabla_v \a)$
\item $\cal{L}_X \o(p; \a) := \o(p; \cal{L}_X(\a)).$
\end{itemize}

\subsection*{Integrals of forms over element chains}    
 If $\dot{P} = \sum a_j \dot{\s}(p_j)$ is a $k$-element chain and $\o$ is a smooth differential $k$-form then 
 the integral $\int_{\dot{P}} \o$ is merely $\sum a_j \o(p_j)(\s(p_j)).$  That is, we simply evaluate the form at each point in the support of $\dot{P}.$  

Stokes' theorem for element chains and smooth forms may now be stated

\begin{theorem} $$\int_{\p \dot{P}} \o = \int_{\dot{P}} d \o.$$
\end{theorem}
The star operator extends directly to $k$-elements $$\star:V_k \to V_{n-k}.$$       Thus we have the star theorem.
 
$$\int_{\star \dot{P}} \o = \int_{\dot{P}} \star \o.$$
Discrete divergence and curl theorems follow.
$$\int_{\p \star \dot{P}} \o = \int_{\star \dot{P}} d \o.$$
$$\int_{\star \p \dot{P}} \o = \int_{\dot{P}} d \star \o.$$
 
\subsection*{Cup product of higher order elements}  Cup   product $\a \cup \b$ is defined for pairs $(\a,\b)$  of  $k$-elements $\a$ of order $s$ and $j$-elements $\b$ of order $t$ , supported in the same point $p$.  The product is a $(k+j)$-element of order $s+t$, also supported in $p$.    We take the $k$-elements $\s$ and $\t$ which generate $\a$ and $\b$, resp., along with the vectors $v_1, \cdots v_s$ and $w_1, \cdots w_t$. The product $\s \cup \t$ is a well defined $(k+j)$-element.  The collection of vectors $v_1, \cdots, v_s, w_1, \cdots, w_t$ produces a differential $(k+j)$-element of order  $(s+t)$.

\begin{theorem}
$$\p (\a \cup \b) = ((\p \a) \cup \b) + (\a \cup (\p \b)).$$
\end{theorem}
\footnote{Sign correction needed.}   

 This definition may be linearly extended to define $A \cup B$ where $A$  and $B$ are higher order element chainlets, supported in the same  finite set of points $J$.

\subsection*{Cup product of chainlets}
It is not possible to define  cup product as a continuous operator
$$\cal{N}^{\natural_{}}_j \times \cal{N}^{\natural_{}}_k \to
\cal{N}^{\natural_{}}_{j+k}$$  that extends cup product of  $k$-elements and coincides with  cup product of differential forms.  Cup product and Hodge star leads to an inner product but     the space of $L^1$ functions, a subspace of chainlets, is not a Hilbert space.   However, cup product is defined on pairs of chainlets where one element of the pair is contained in a dense subspace of chainlets,  namely the exterior chainlets defined earlier.
  $$\cal{N}^{\natural_{}}_k \times \cal{E}^{\natural_{}}_j \to
\cal{N}^{\natural_{}}_{j+k}.$$  
     
     Cup product converges in the natural norm, to define $A \cup Ch(\o)$ where $A$ is an arbitrary chainlet and $Ch(\o)$ is an exterior chainlet.    This extends the standard wedge product on forms  $Ch(\eta \wedge \o) = Ch(\eta) \cup Ch(\o)$.   This leads to an inner product  $$<A,Ch(\o)> = \int_{A \cup *Ch(\o)} dv.$$

We first define cup product for pairs of element chains and exterior chainlets and then prove continuity in the first variable.  

Let $Ch(\o)$ be the exterior $j$-chainlet associated to a differential $j$-form $\o$.  In local coordinates,  $\o(p) = \sum a^H(p) dx^H.$   The differential $j$-form $dx^H$  at $p$ determines a unique unit $j$-element $Ch(e^H)$ supported in $p$ satisfying $dx^H(Ch(e^H)) = 1.$ 
 
Define $$Ch(\o)(p) := \sum a^H(p) Ch(e^H).$$    
\begin{lemma}
Let $p,q \in \R^n.$   If $\o$ is smooth then 
 $$|Ch(\o)(p) - Ch(\o)(q)|^{\natural_{}} \le |p-q||\o|^{\natural_{}}.$$
\end{lemma} 

Let $\dot{\s}$ be a  $k$-element supported in $p$.   
Cup product of $k$-vectors and $j$-vectors is defined for $k$-elements and $j$-cells supported in a point $p$, of course, since it is defined for all vector spaces.  This can be used to define   $$\dot{\s} \cup Ch(\o)(p).$$   If $\dot{P} = \sum a_i \dot{\s}_i$ is a  $k$-element  define 
$$\dot{P} \cup Ch(\o) = \sum a_i \dot{\s}_i \cup Ch(\o).$$ 
\begin{theorem}
$$|\dot{P} \cup Ch(\o)|^{\natural_{}} \le |\dot{P}|^{\natural_{}} |\o|^{\natural_{}}.$$
\end{theorem}

\begin{proof}

Suppose $\dot{P}$ is a $k$-element   and $\e > 0$.  There exists a decomposition $\dot{P} = \sum \dot{D}^i $ with $$|\dot{P}|^{\natural_{}} > \sum \|\dot{D}^i\|_i   - \e.$$   Then 
$$|\dot{P} \cup Ch(\o)|^{\natural_{}} \le \sum 
|a_i||\dot{D}^i \cup Ch(\o)|^{\natural_{}}.    $$

This reduces to showing $\|\dot{D}^i \cup Ch(\o)\|_i \le \|\dot{D}^i\|_i |\o|_i$
   
Now $\dot{\s}^i \cup Ch(\o)$ can be written as the sum of two chainlets, the first is an $i^{th}$ order dipole based on $\dot{\s} \cup Ch(\o)$ and vectors $v_1, \dots, v_i.$  Its dipole norm is bounded by $$\|\dot{\s} \cup Ch(\o)\|_0|v_1| \cdots |v_i| \le \|\dot{\s}\|_0|v_1| \cdots |v_i| |\o|_0 =  \|\dot{\s}^i\|_i  |\o|_0 .$$  The second term is bounded by $$\begin{aligned} M(\s)(|\o(p) - \o(p+v_1)  &+ \o(p+v_1+v_2) -\o(p+v_2)| + \cdots) \\&\le M(\dot{\s})|v_1||v_2| \cdots |v_i| |\o|_i \\&= \|\dot{\s}^i\|_i|\o|_i.\end{aligned}$$  Hence 
$$\|\dot{\s}^i \cup Ch(\o)\|_i \le \|\dot{\s}^i\|_i |\o|_i.$$  
 
In the top dimensional case, $k = n$, the result follows since $\dot{P} \cup Ch(\o) = 0.$ 
By induction, 
\[
\begin{aligned} \sum 
|a_i||\dot{D}^i \cup Ch(\o)|^{\natural_{}}   &\le \sum |a_i| \|\dot{D}^i \cup Ch(\o)\|_i    \\&\le 2|\o|^{\natural_{}}\sum |a_i|\|\dot{D}^i\|_i   \\& \le 2|\o|^{\natural_{}}\left(\sum |a_i|\|\dot{D}^i\|_i  \right) 
\end{aligned}
\]
\end{proof} 
  
\subsection*{Properties of cup product} 
\begin{enumerate}
\item Associative
\item Bilinear   $$(A+B) \cup Ch(\o) = (A \cup Ch(\o)) + (B \cup Ch(\o)).$$
  $$A \cup (Ch(\o) + Ch(\eta)) = (A \cup Ch(\o)) + (A \cup Ch(\eta)).$$
\item Operators:  $$\p ( A \cup Ch(\o) ) = (\p A) \cup Ch(\o) + (A \cup \p Ch(\o)).$$
\item $$\star  (A \cup Ch(\o) ) = \star A \cup Ch(\star \o)$$
\item $$f_*( A \cup Ch(\o) ) =   f_* A \cup f_*(Ch(\o))$$
\item $$(X \cap M) \cup (Y \cap M) = \int_M \phi(X) \cup \star \phi(Y) = (X \cup Y) \cap M$$
 \item anti commutative  $$A \cup \star B = (-1)^k B \cup \star A$$   
\item $A \cup A = 0$ if $A$ is simple.  
\end{enumerate}

\begin{theorem} If $A$ is a   $k$-chainlet of class $N^r$ and $\o$ is a $k$-form of class $B^r$ then
$$\int_A \o = \int_{A \cup \star Ch(\o)} dV.$$ 
\end{theorem}
  
 \begin{corollary} Assume $M$ has no boundary.  Then  $\d  \o = 0 \iff  \p Ch(\o) = 0.$ Also, $d \o = 0 \iff  \diamondsuit Ch(\o) = 0.$
$\o$ is harmonic $\iff$ if $Ch(\o)$ is harmonic.
\end{corollary}  

\begin{proof} By the Leibnitz rule for wedge product if $\a$ is a $p$-form and $\b$ a $q$-form then, $d(\a \wedge \b) =  (-1)^p \a \wedge d \b + d \a \wedge \b.$  

Assume  $\d \o = 0,$  Then $d \star \o = 0$.   Then $$X \cdot \p Ch(\o) = dX \cdot Ch(\o) = \int d\phi(X) \wedge \star \o =(-1)^{k+1}\int \phi(X) \wedge d \star \o = 0.$$  Since this holds for all $X$ we deduce $ \p Ch(\o) = 0.$
Similarly, $$\d(\a \wedge \b) = \star d \star (\a \wedge \b) = \star d(\star \a \wedge \star \b) = \star (d \star \a \wedge \star \b + \star \a \wedge d \star \b) = \d \a \wedge \b + \a \wedge \d \b.$$    Note that $\diamondsuit M = 0$.  Hence $\int_M \d(\phi(X) \wedge \star \o) = 0.$   

Assume $d \o = 0$.  Then $\d \star \o = 0.$    Let $\o$ be a $(k-1)$-form and $X$ a $k$-cochain.  Then $\d(\phi(X) \wedge \star \o) = 0.$  Then $$X \cdot \diamondsuit Ch(\o) = \d X  \cdot Ch(\o) = \int \d \phi(X) \wedge \star \o = \int \phi(X) \wedge \d \star \o = 0.$$  It follows that $ \diamondsuit Ch(\o) = 0.$

Conversely, suppose $\p Ch(\o) = 0. $ Then $$\d \o \cdot P = \o \cdot \diamondsuit P = \int_{\diamondsuit P \cup \star Ch(\o)} dV = \int_{P \cup \diamondsuit \star Ch(\o)} dV = 0.$$   Here, we use $\d dV = 0.$    Now suppose $  \diamondsuit Ch(\o) = 0.$  Then 
$$d \o \cdot P = \o \cdot \p P = \int_{\p P \cup Ch(\o)} dV = \int_{P \cup \p Ch(\o)} dV  = 0.$$ Here we use  $\int_{\p (P \cup Ch(\o))}dV = 0.$
\end{proof}

\begin{theorem}[Cartan's magic formula for differential elements] \label{Cartan}
$$\cal{L}_X(\a) = (Ch(\phi(X)) \cup \p \a) + \p (Ch(\phi( X)) \cup \a).$$
\end{theorem}

Define $i_X \a := Ch(\phi(X)) \cup \a.$  Then 
$i_X \o(p; \a) = \o(p; i_X \a).$
\begin{corollary}[Cartan's magic formula for forms]
$$\cal{L}_X \o =d(i_X \o) + i_X d \o .$$
\end{corollary}

\begin{proof} By Theorem \ref{Cartan}
$$\begin{aligned}\cal{L}_X \o(p; \a) &= \o(p; \cal{L}_X(\a)) \\&= \o(p; Ch(\phi(X)) \cup \p \a) + \o(p; \p (Ch(\phi(X)) \cup \a)) \\&= d(i_X \o)(p; \a) + i_X d \o(p;\a).\end{aligned}$$
\end{proof}
  
There is an orthonormal basis to $k$-element chains supported in a finite set of points $J$. An inner product can be defined on these chains leading to  new numerical methods.  
 
\


\begin{thebibliography}{rll}
\bibitem{plateau} J.A.F. Plateau, Statique Experimentale et Theorique des Diquides Soumis aux Seules
Forces Moleculaires, Paris, Gauthier-Villars, 1873
\bibitem{whitney} Whitney, Hassler, Geometric Integration Theory, 1959
\bibitem{whitney2} Whitney, Hassler, Moscow 1935: Topology Moving Toward America, Collected Papers, Vol 1, Birkhauser, 1992.
\bibitem{federer} Federer, Herbert, {\sc Geometric Measure Theory},
Springer-Verlag, New York, 1969
\bibitem{weyl}  Weyl, Herman, {\em Philosophy of 
Mathematics and Natural Science}, Princeton University Press, 1949
 
\bibitem{stokes} Harrison, Jenny, {\em  Stokes'
 theorem on nonsmooth chains}, Bulletin AMS, October
 1993.
\bibitem{continuity} --, Continuity of the Integral as a Function of the Domain, Journal of Geometric Analysis, 8 (1998), no. 5, 769--795
\bibitem{iso} --.Isomorphisms differential forms and cochains, Journal of Geometric Analysis, 8
(1998), no. 5, 797--807.
 \bibitem{currents} --, {\em  Geometric
 realizations of currents and distributions},  Proceedings of Fractals and Stochastics III, Friedrichsroda, German, 2004.
\bibitem{hodge} --,Geometric Hodge star operator with applications to the theorems of Gauss and Green, to appear, Proc Cam Phil Soc.
\bibitem[H4]{madeira} --, Flux across nonsmooth boundaries and fractal Gauss/Green/Stokes theorems, J. Phys. A 32 (1999), no. 28, 5317--5327.
8, 5317--5327.
Flux across nonsmooth boundaries and fractal Gauss/Green/Stokes theorems, J. Phys. A 32 (1999), no. 28, 5317--5327.
 
 
 
 \bibitem{flanders} Flanders, Harley, Differential forms with applications to the physical science, Dover.
 
\bibitem{feich} Feichtinger, Hans, {\em Compactness in translation invariant Banach spaces of distributions and compact multipliers},   J. Math. Anal. Appl. 102, 1984, 289Ð327. 
\end{thebibliography}
\end{document}